\documentclass[12pt,preprint]{aastex}

\lefthead{P. Candia {\em et al.}}
\righthead{Supernova 2000cx}

\begin{document} 

\title{Optical and Infrared Photometry of the Unusual Type Ia 
Supernova  2000cx}
\shorttitle{SN 2000cx}
\shortauthors{P. Candia et al.}

\author{P. Candia, K. Krisciunas, Nicholas B. Suntzeff, D. Gonz\'{a}lez,
J. Espinoza, R. Leiton, A. Rest, and R. C. Smith}

\affil{Cerro Tololo Inter-American Observatory, National Optical Astronomy
Observatories,\altaffilmark{1} Casilla 603, La Serena, Chile \\
pcandia@ctio.noao.edu, kkrisciunas@noao.edu, nsuntzeff@noao.edu,
leiton@ctio.noao.edu, arest@noao.edu, csmith@noao.edu}

\author{J. Cuadra}
\affil{Pontificia Universidad Cat\'{o}lica de Chile, Santiago, Chile \\
jcuadra@puc.cl}

\author{T. Tavenner}
\affil{New Mexico State University, Astronomy Department, Box 30001, MSC 4500, Las
Cruces, NM 88003-8001 \\
tanya@nmsu.edu}

\author{C. Logan}
\affil{Department of Earth, Atmospheric, and Planetary Sciences, Massachusetts 
Institute of Technology, Cambridge, MA 02139 \\
clogan@mit.edu}

\author{K. Snider, M. Thomas, and A. A. West}
\affil{Department of Astronomy, University of Washington, Box 351580, Seattle,
WA 98195 \\
ksnider@u.washington.edu, drtas@u.washington.edu, west@astro.washington.edu}

\author{G. Gonz\'{a}lez}
\affil{Iowa State University, Deparment of Physics and Astronomy, Ames, IA 50011-3160 \\
gonzog@iastate.edu}

\author{S. Gonz\'{a}lez and M. M. Phillips}
\affil{Las Campanas Observatory, Casilla 601, La Serena, Chile \\
sergiogonzalezctio@yahoo.com,mmp@lco.cl}

\author{N. C. Hastings and R. McMillan}
\affil{Apache Point Observatory, Astrophysical Research Consortium, 2001 Apache
Point Road, P. O. Box 59, Sunspot, NM 88349-0059 \\
hastings@apo.nmsu.edu, mcmillan@apo.nmsu.edu }

\altaffiltext{1}{The National Optical Astronomy
Observatories are operated by the Association of Universities for
Research in Astronomy, Inc., under cooperative agreement with the
National Science Foundation.}

\begin{abstract}

We present optical and infrared photometry of the unusual Type Ia
supernova 2000cx. With the data of Li et al. (2001) and Jha
(2002), this comprises the largest dataset ever assembled for
a Type Ia SN, more than 600 points in $UBVRIJHK$.  We confirm the finding
of Li et al. regarding the unusually blue $B-V$ colors as SN 2000cx
entered the nebular phase.  Its $I$-band secondary hump was extremely
weak given its $B$-band decline rate.  The $V$ {\em minus} near infrared
colors likewise do not match loci based on other slowly declining
Type Ia SNe, though $V-K$ is the least ``abnormal''.  In several ways
SN 2000cx resembles other slow decliners, given its $B$-band {\em decline}
rate ($\Delta$m$_{15}$($B$) = 0.93), the appearance of 
Fe III lines and weakness of Si II in its pre-maximum spectrum,
the $V-K$ colors and post-maximum $V-H$ colors.  
If the distance modulus derived from Surface Brightness Fluctuations
of the host galaxy is correct, we find that the rate of light
increase prior to maximum, the characteristics of the bolometric light
curve, and the implied absolute magnitude at maximum are all
consistent with a sub-luminous object with $\Delta$m$_{15}$($B$)
$\approx$ 1.6-1.7 having a higher than normal kinetic energy.

\end{abstract}

\parindent = 0 mm

Keywords: (stars:) supernovae: individual, SN~2000cx, SN~1986G, 
SN~1999aa, SN~1999ac, SN~1999aw, SN~1999ee, SN~1999gp, SN~2000bk, SN~2001ba,
SN~2001el; infrared: stars; techniques: photometric

\section{Introduction}

\parindent = 9 mm

Astronomers try to understand the universe by looking for patterns in
observed phenomena.  Often, the patterns themselves are reason to believe in
underlying, understandable physical mechanisms, while at other times the
exceptions to the rules provide the motivation to expand our conceptions
of the physical makeup of cosmic objects. In this paper we present
optical and infrared photometry of the very unusual supernova 2000cx.  
Previous optical data have been presented by Li et al. (2001), who
describe SN 2000cx as ``unique''.
%\footnote[2] {In an old Monty Python
%skit John Cleese plays a woman named Anne Elk, who is on a television
%talk show to discuss her theory of the Brontosaurus.  She says, ``All
%Brontosauruses are thin at one end, much, much thicker in the middle, and
%then thin again at the far end.'' Well, SN 2000cx blew up, got much, much
%brighter in the middle, and faded away at the other end.  {\em How} it
%got brighter and then fainter again was unlike any Type Ia SN that has
%been studied so far.}

SN 2000cx was discovered by Yu, Modjaz, \& Li (2000) from images taken on
17.5 and 18.4 July 2000 UT as part of the Lick Observatory Supernova Search,
using the 0.76-m Katzman Automatic Imaging Telescope (KAIT).  This object was
located at $\alpha$ = 1:24:46.15, $\delta$ = +9\arcdeg ~30\arcmin ~30\farcs9
(equinox 2000.0), which is 23\farcs0 west and 109\farcs3 south of the
nucleus of the S0 galaxy NGC 524. A spectrum taken on 23 July UT with the
Nickel 1-m reflector at Lick Observatory (Chornock et al. 2000) revealed
that the object was a peculiar Type Ia supernova, resembling 
SN 1991T a few days before maximum brightness, with prominent
Fe III absorption lines near 430 and 490 nm but weak Si II at 612 nm.
Optical photometry (Li et al. 2001)  revealed that SN 2000cx is different
from all known Type Ia SNe and that the light curves cannot be fitted well
using the techniques currently available.  The pre-maximum rise was
relatively fast, similar to SN 1994D, but the post-maximum decline was
relatively slow, similar to SN 1991T.

We present optical and infrared photometry of SN 2000cx, initiated at
CTIO with the Yale-AURA-Lisbon-Ohio (YALO) 1-m telescope on 19 July 2000
(UT), some 8 days before the time of $B$-band maximum.  We include data
taken with the 0.76-m Manastash Ridge Observatory (MRO) of the University
of Washington (also begun on 19 July UT), the CTIO 0.9-m telescope, and
the Apache Point Observatory 3.5-m telescope (APO). The calibration of
the infrared light curves was primarily made possible with observations
made with the Swope 1-m telescope at Las Campanas Observatory.

\section{Observations}

The YALO optical images were obtained using a Loral 2048 $\times$ 2048
CCD with a scale of 0\farcs30 per pixel, giving roughly a 10\arcmin
$\times$ 10\arcmin ~field of view. Due to amplifier problems only half of
the chip was working before September 2000. On 6 September 2000 this
problem was fixed, but it was necessary to change the gain from 6.6 to 3.6
electrons per ADU.  The readnoise improved from 14.2 to 11 electrons.
The broadband $JHK$ imges from the YALO
telescope were obtained using a 1024 $\times$ 1024 HgCdTe {\sc hawaii}
Array from Rockwell. The filters have over an 80 percent transmittance
from 1.171 to 1.322 $\mu$m for the $J$-band, 1.498 to 1.766 $\mu$m for
$H$ and 2.012 to 2.278 $\mu$m for $K$. The scale of the infrared CCD is
0\farcs20 per pixel, making a total field of view of 3\farcm3 $\times$
3\farcm3.

The camera used at the CTIO 0.9-m telescope contains the \#3
Tektronix 2048 CCD, which is a thinned, anti-reflection coated, 
back-side illuminated chip with 2K $\times$ 2K pixels.  
The scale and
field size are 0\farcs40 per pixel and 13\farcm5, respectively. The
filters used were: 
$U$ (liquid CuSO$_4$), $\lambda_{0}$ =  3575 \AA, {\sc fwhm} =
600 ~\AA; 
$B$, $\lambda_{0}$ = 4202 ~\AA, {\sc fwhm} = 1050 ~\AA; 
$V$, $\lambda_{0}$ = 5475 ~\AA, {\sc fwhm} = 1000 ~\AA;
$R$, $\lambda_{0}$ = 6425 ~\AA, {\sc fwhm} = 1500 ~\AA;
$I_{KC}$, $\lambda_{0}$ = 8075 ~\AA, {\sc fwhm} = 1500 ~\AA. 
The $UBVRI_{KC}$ filters were from the CTIO facility set Tek \#1.  

Traces of the filters
used with the YALO and CTIO 0.9-m telescopes are shown in Stritzinger
et al. (2002).  However, for our SN~2000cx observations with YALO
we did {\em not} use the very broad, non-standard $R$ filter used
by Stritzinger et al.  Instead, we used a much narrower, more standard
filter (see Appendix).

The APO 3.5-m optical images were obtained using the facility CCD
imager SPIcam, which contains a back-side illuminated SITe chip of 2048
$\times$ 2048 pixels.  2 $\times$ 2 readout was used, giving a scale
of 0\farcs28 per pixel and a 4\farcm78 $\times$ 4\farcm78 field of
view. The APO infrared images were obtained with the 3.5-m telescope 
using GRIM II, which contains a 
Rockwell 256 $\times$ 256 {\sc nicmos}3 HgCdTe array.  The chip is sensitive
from 1 to 2.5 $\mu$m with a quantum efficiency of approximately 70 percent,
a gain of 4.7 electrons per ADU, and a readnoise of 110 electrons.
The $J$, $H$ and $K^{\prime}$ filters transmit at 1.265 $\pm$ 0.267,
1.646 $\pm$ 0.339, and 2.114 $\pm$ 0.343 $\mu$m, respectively.

The CCD camera at the MRO 0.76-m telescope uses a Ford
Aerospace chip of 1024 $\times$ 1024 pixels, with a readnoise of 8 electrons.  
The scale is 0\farcs61 per pixel, giving a 10\farcm32 $\times$ 10\farcm32 
field of view. It contains $UBVRI$ ``Harris'' filters.

Instrumental YALO and 0.9-m magnitudes were measured as
Point Spread
Function (PSF) magnitudes using the {\sc daophot} package (Stetson 1987,
1990).  A transformation equation of the form $m$ = f(M,I,X,T), as suggested
by Harris, Fitzgerald, \& Reed (1981), was used. Here $m$ is the observed (i.e.
instrumental) magnitude, M is the tabulated magnitude (e.g. from
Landolt 1992), I is the tabulated color index, X is the 
airmass, and T is the time during the night.

Using the observations of the Landolt (1992) standards, we determined
zeropoints, color terms and atmospheric terms.  This allowed us to
calibrate the field stars near SN 2000cx.  Photometry of the supernova
itself is then tied to the Landolt standards via these field stars.  This
allows us to derive accurate values for the SN even if it is observed
under non-photometric conditions.  Once the field star sequences were
established, we dropped the airmass term because the differential airmass
corrections within a CCD frame are negligible.  The time-variable term was
dropped because of its demonstrated small contribution to the final
photometry (Suntzeff et al. 1999).

Once the MRO and APO images were bias subtracted and flattened, we
obtained aperture magnitudes in the {\sc iraf} environment, using {\bf
phot} within the {\bf apphot} package, and calibrated the field stars
using {\bf mknobsfile}, {\bf
fitparams}, and {\bf evalfit} within the {\bf photcal} package.  The
transformation equations were configured to produce $V$ magnitudes,
$B-V$, $V-R$, and $V-I$ colors. The APO infrared data were reduced using
{\sc iraf} scripts written by Alan Diercks, and the IR mosaics were
produced using Eugene Magnier's image processing program {\bf mana}.
We then carried out aperture photometry within {\sc iraf}.

The mean optical transformation coefficients for the four optical systems
described above are to be found in Table 1.
 
Figure 1 shows a combined $BVR$ image taken with the CTIO 0.9-m telescope
when SN 2000cx was on the rise.  Some nearby field stars are marked.
Optical photometry of these stars, obtained from imagery with YALO and MRO,
is to be found in Table 2.  A comparison of the independent YALO, MRO,
and Li et al. (2001) values for the field stars marked as numbers 2, 4,
and 8 in Fig. 1 shows good agreement.  The {\em range} of the  mean values 
for these three stars is 16-29 mmag in $B$, 12-17 mmag in $V$,
14-30 mmag in $R$, and 6-36 mmag in $I$, respectively.
Thus, photometry of {\em stars} can be carried out at the $\sim$0\fm02 mag level.

Absolute calibration in $J$, $H$, and $K$ was done via observations of
the SN 2000cx field along with infrared standards of Persson et al.
(1998), using the LCO Swope 1-m telescope on five nights in October and
November 2001. The mean $JHK$ values for field stars 1 and 2 are given in
Table 2. Because of the size of the field of view of the
IR camera on the APO 3.5-m telescope, only star 1 and SN 2000cx were
always on the chip while dithering.  To make the fullest use of all the
nights when IR data were taken (i.e. photometric and non-photometric
nights), we chose to reduce all our IR photometry of SN 2000cx (i.e. from
YALO {\em and} APO) with respect to field star 1.

Our infrared YALO images were reduced using a package of scripts written by one of us
(NBS),\footnote[2]{http://www.ctio.noao.edu/~nick/reduction/reduction.html}
which runs in the {\sc iraf} environment. This package contain tasks which fill
out the file headers with information necessary for subsequent reduction and
take care of bias correction, field flattening, masking out bad pixels, and
vignetting.  Fortunately, field star number 1 and the SN were located out of the
zone of vignetting so that we did not have to worry about that at all. A section
free of stars was chosen for each night so we could calculate a clean sky level
and subtract that from all individual frames.

Tables 3 through 6 contain our optical data from the CTIO 0.9-m, YALO, APO,
and MRO telescopes, respectively.  In effect, we have four independent
optical datasets, with two independent calibrations. The uncertainties of the
values in the tables derive from photon statistics and the uncertainties of
color corrections, and are to be considered minimum internal errors.  A more
accurate estimate of the accuracy of our photometry is obtained by fitting
fourth order polynomials to the light curves, telescope by telescope and 
filter by filter, to
measure the {\sc rms} residuals of such fits, under the assumption that the
variation of light of the SN is a smooth function for each filter. We have calculated
these residuals for our two largest datasets, using data prior to 25 d
after the time of $B$-band maximum.  
For the YALO data the internal
errors are $\sigma _B = \pm$ 14; $\sigma _V = \pm$ 34; $\sigma _R = 
\pm$ 41; and $\sigma_I = \pm$ 71 mmag.  For the MRO data the internal
errors are $\sigma _B = \pm$ 6; $\sigma _V = \pm$ 20; $\sigma _R = 
\pm$ 43; and $\sigma_I = \pm$ 51 mmag.  

%kkk

Figure 2 shows our optical data. Fig. 3 is the
same as Fig. 2, but with the addition of the KAIT and Wise Observatory
data given by Li et al. (2001), plus data obtained by Jha (2002)
with the 1.2-m telescope at the Fred L. Whipple Observatory at Mt.  Hopkins,
Arizona.

There are systematic differences between different subsets of the data.
With the sense of $\Delta \equiv$ ``YALO {\em minus} MRO'', at the time
of $B$-band maximum $\Delta B = -$0.04, $\Delta V = -$0.05, $\Delta R = -$0.04,
$\Delta I$ = +0.05 mag.  At $t$ = 17.4 d, from Tables 4 and 6 we find 
$\Delta B = -$0.02, $\Delta V = -$0.08, $\Delta R = -$0.15, $\Delta I$ = +0.16 mag.
The disagreement in $V$, $R$, and $I$ is large.

Now, with the sense of $\Delta \equiv$ ``CTIO 0.9-m {\em minus} MRO'', for
the data within 7 days of maximum we find $\Delta B = -$0.02, $\Delta V =
-$0.04, $\Delta R = -$0.05, $\Delta I$ = 0.00 mag.  This is not a
significant improvement on the YALO vs. MRO situation {\em at maximum}.  
However, YALO photometry vs. CTIO 0.9-m data within a week of maximum is
in agreement at the 0\fm02 level or better for $B$, $V$, and $R$.  In
spite of systematic differences between datasets obtained with different
telescopes and different filters, at least we know the apparent magnitudes
at maximum to $\pm$ 0.03 mag.

The systematic differences in the photometry
are undoubtedly due to differences in the actual filters used, coupled
with the non-stellar spectral energy distribution of the SN.
Very late-time $I$-band data are particularly
discordant, more than 0.7 mag (KAIT vs. FLWO).
Though some of the optical spectra discussed by Li et al.
(2001) cover the full $I$-band (at 2, 6, 7, 32, 42 days after
T($B_{max}$) and later), to reconcile all data of SN 2000cx
taken with different telescopes
is beyond the scope of this paper.\footnote[3]{Stritzinger et al. (2002)
attempted to correct data of SN 1999ee obtained with multiple telescopes
by calculating ``S-corrections'' using the spectra of this object plus
knowledge of the filter transmission curves, quantum efficiencies of the
chips, and appropriate atmospheric transmission functions.  In the end
they decided to leave their data uncorrected.  However, more recently
Krisciunas et al. (2003) derived corrections to their photometry of SN
2001el and applied the filter corrections to the $BVJHK$ data.  This in
particular solved problems with the $B$-, $V$-, $J$-, and $H$-band data.
However, they found that $R$-band corrections were not really necessary,
and that applying $I$-band corrections actually made datasets obtained
with different telescopes much {\em more} discordant.}

Tables 7 and 8 contain the infrared data from YALO and APO, respectively.
Table 9 has corrections to place the APO data (which are fewer in number)
on the YALO filter system, using the method of Stritzinger et al. (2002)
and Krisciunas et al. (2003).  To our knowledge, only two infrared
spectra of SN 2000cx itself exist (Rudy et al. 2002), and these were
taken 6 and 7 days, respectively, before the observed date of $B$-band
maximum.  To calculate the filter corrections for the APO data of
SN 2000cx we used the infrared spectra of SN 1999ee (Hamuy et al. 2002).
In Figs. 4 and 5 we show our infrared photometry without, and with, the
filter corrections.  The agreement of the $J$-band datasets is clearly
better with the corrections.

Finally, in Table 10 we give the times of maximum, as observed by
us in the different bands.  Our time of $B$-band maximum is 0.3~d
later than that found by Li et al. (2001), well within the uncertainty.
While they found that the $V$-band maximum occurred 2.1~d
after $B$-band maximum, we find a lag of only 1.2~d.  However, this
difference is not statistically significant.

Jha (2002) also obtained $UBVRI$ photometry of SN 2000cx, though with a
large gap in time after maximum.  Late time photometry (10 July 2001)
with $HST$, using the $F675W$ and $F814W$ filters, is given by Li et al.
(2002). Altogether, the photometric database of SN 2000cx amounts to 642
data points.  To our knowledge it is the largest dataset ever obtained
for a Type Ia SN.

\section{General Discussion}

A vast majority of the light curves of Type Ia SNe can be fit into a
classification scheme in which the shape of the light curves is correlated
with the intrinsic luminosity of the SN (Phillips 1993, Riess, Press, \&
Kirshner 1996, Perlmutter et al. 1997, Phillips et al. 1999).  Li et al.
(2001) remark that the Multi-color Light Curve Shape (MLCS) fit to their
data for SN 2000cx is the worst fit they have ever seen. SN 2000cx was a
reasonably fast riser, but a slow decliner.  That would mean that the
``stretch factor'' used with the method of Perlmutter et al. (1997) would
give a different value prior to maximum compared to after maximum.  Since
the stretch factor in $B$ and $V$ is related to the intrinsic luminosity,
how does one determine the intrinsic luminosity of this SN?

NGC 524, the host galaxy of SN 2000cx, has had its distance modulus
measured via the Surface Brightness Fluctuation (SBF) method.  Tonry
et al. (2001) obtain $m-M$ = 31.90 $\pm$ 0.20 for the distance modulus. Li 
et al. (2001) obtained $m-M$ = 32.53 $\pm$ 0.35 using MLCS.  Using the corrected
recession velocity of NGC 524 in the Local Group frame of 2192 km
s$^{-1}$ and a Hubble constant of 74 km s$^{-1}$ Mpc$^{-1}$, in agreement with   
the $HST$ Key Project value (Freedman et al. 2001), we obtain a distance 
modulus of 32.36 mag.

SN 2000cx was located in the outer regions of an early-type galaxy.  Since
early-type galaxies contain minimal amounts of dust, and the $B-V$ colors
of SN~2000cx were particularly blue, we explicitly assume that this
SN was unreddened in its host.  We shall correct for reddening due to
dust in our Galaxy (see below).

Krisciunas et al. (2003) showed that the $H$-band absolute magnitudes
of Type Ia SNe 10 days after the time of $B$-band maximum appear to be a
flat function of the decline rate $\Delta$m$_{15}$($B$).  For a sample of
9 objects with $\Delta$m$_{15}$($B$) $<$ 1.3
they found a mean $H$-band absolute magnitude of $-$17.91
$\pm$ 0.05.  The observed $H$-band magnitude of SN 2000cx 10 days
after $B$-band maximum is 14.64 $\pm$ 0.04.  From the Galactic reddening maps
of Schlegel, Finkbeiner, \& Davis (1998), we estimate that the color
excess E($B-V$) = 0.082 mag in the direction of SN 2000cx. Using the
interstellar extinction model of Rieke \& Lebofsky (1985), we
estimate that the $H$-band extinction, due only to the effect of
dust in our Galaxy, is 0.04 mag.  Assuming that the absolute magnitude
of SN 2000cx is also $-$17.91, given its corrected $H$-band
magnitude of 14.60, we obtain a distance modulus $m-M$ = 32.51.  
Thus, the distance moduli from MLCS, Hubble's Law, and the $H$-band
analysis are in reasonable agreement, with an unweighted mean value
of 32.47 $\pm$ 0.05, corresponding to a distance of 31 $\pm$ 1 Mpc.  

Given the observed $V$-band maximum of 13.25, A$_V \approx$ 
3.1 $\times$ 0.082 = 0.25, and a distance modulus of 32.47, the
implied $V$-band absolute magnitude of SN 2000cx is $-$19.47, which is
comparable to the mean of M$_V$ of the slow decliners (Krisciunas
et al. 2003, Fig. 13).

However, Ajhar et al. (2001) have shown that there is excellent agreement
between distance determinations using SBF and other methods.  The host of
SN 2000cx is actually a bit close ($cz < 3000$ km s$^{-1}$) to derive its
distance via Hubble's
Law.  The MLCS distance can also be doubted because the light curves
cannot be fit well using MLCS templates.  Also, on the basis of
a larger sample of objects, it may turn out that $H$-band absolute magnitudes
do show some kind of decline rate relation.  In that case, should we use
the {\em decline} rate of SN 2000cx or some modified value that takes into
account its different rise and decline rates?

Let us assume that the Tonry et al. (2001) distance modulus of 
$m-M$ = 31.90 $\pm$ 0.20 is correct.  On an $H_0$ = 74 scale this would
be $m-M$ = 31.84 mag. Adopting A$_B$ = 0.34, A$_V$ = 0.25,
and A$_I$ = 0.12 mag and the maximum magnitude values given in Table 10, we
obtain absolute magnitudes of M$_B$ = $-$18.76,
M$_V$ = $-$18.84, M$_I$ = $-$18.31, with uncertainties of $\pm$ 0.20 mag.
These correspond to $\Delta$m$_{15}$($B$) in the range 1.4-1.7 (Krisciunas
et al. 2003, Fig. 13).  M$_H$ ($t$ = 10 d) would be $-$17.24, comparable
to that of SN~2000bk, which was a fast decliner with $\Delta$m$_{15}$($B$)
= 1.63.  

Krisciunas et al. (2001, Figs. 16 and 17) devised a quantitative measure
of the strength of the $I$-band secondary hump common to Type Ia SNe,
namely the mean flux-with-repect-to-maximum from 20 to 40 d after the
time of $B$-band maximum.  They found that 90 percent of Type Ia SNe
have values of  $\langle$ I $\rangle _{20-40}$ that are well correlated
with the $B$-band decline rate $\Delta$m$_{15}$($B$).  There were two
exceptions to the rule, showing that there can be objects with identical
decline rates in $B$ and $V$ but much stronger or weaker secondary humps
in $I$.  For SN 2000cx we find that  $\langle$ I $\rangle _{20-40}$ =
0.35.  This would imply $\Delta$m$_{15}$($B$) $\approx$ 1.7 like
the fast decliners SNe 1992bo or 1993H.  But SN 2000cx is a slow
decliner, with $\Delta$m$_{15}$($B$) = 0.93.  If we were to add a point
in Fig. 17 of Krisciunas et al. (2001) corresponding to SN 2000cx,
it would be the most discrepant point in the graph.

Li et al. (2001) point out that the stretch factor for the pre-maximum
data points ($t = -8$ to 1 day) corresponds to $\Delta$m$_{15}$($B$) = 
1.64 $\pm$ 0.02.  In various ways SN~2000cx masquerades as a fast
Type Ia SN and also as a slow one.

One of the patterns exhibited by many unreddened Type Ia SNe is that their
$B-V$ colors from 30 to 90 d after the time of $V$-band maximum follow a
particular linear trend, whether they are fast-decliners or slow-decliners
(Lira 1995, Phillips et al. 1999).  Li et al. (2001) and Cuadra et al.
(2001) noted that SN 2000cx was roughly 0.2 mag bluer than the Lira
line.  In Fig. 6 we show the $B-V$ data from YALO, the CTIO 0.9-m,
APO, and MRO telescopes.  The data have been dereddened by 0.082 mag
to account for dust in our Galaxy (Schlegel et al. 1998).  The data
after JD 2,451,784 are, on average, 0.217 mag below the Lira line.
We assume, because SN~2000cx occurred in the outer regions of an
early-type galaxy, that it suffered no host reddening.  But if its light
was affected by dust in the host galaxy, then the points in Fig. 6 should
be displaced even further below the Lira line.  Truly, the $B-V$ colors of
this SN are unusual.

In Figs. 7, 8, and 9 we show the $JHK$ light curves of SN 2000cx
along with data of other objects: SNe 1999aw (Strolger et al. 2002),
2001el (Krisciunas et al. 2003), 1999ac (Phillips et al. 2002, 2003),
2000bk (Krisciunas et al. 2001), and 1986G (Frogel et al. 1987).  The
light curves are ordered from top to bottom by the decline rate
$\Delta$m$_{15}$($B$).  

We the note the extremely deep dip in the $J$-band in the 20 days
after maximum. We also note the similarity of the SN~2000cx $J$-band data
with that of SN~2000bk, a fast decliner.  The latter had an earlier
secondary $J$-band maximum, however.
We also note, whereas most Type Ia SNe have relatively flat
$H$- and $K$-band light curves in the 20 days after maximum, SN 2000cx
showed decreasing flux at this epoch.

Krisciunas et al. (2000, 2001, 2003) found that Type Ia SNe which are {\em
mid-range} decliners appear to exhibit uniform $V minus$ infrared
color curves from about one
week before maximum until three or more weeks after maximum.  They asserted
that the unreddened loci can be used to derive the total extinction
suffered by the light of a Type Ia SN on its way to our viewing location in
the Galaxy.  They found that $V-H$ and $V-K$ were particularly well
behaved, and modelling by H\"{o}flich given in Krisciunas et al. (2003)
confirms this observational result from a theoretical standpoint.
Krisciunas et al. (2000, 2001) also noted that fast decliners and slow
decliners have different unreddened loci.  The slow decliners have bluer
loci, and the fast decliners have redder loci.

Using data of the slow decliners SNe 1999aa (Krisciunas et al. 2000),
1999aw (Strolger et al. 2002), and 1999gp (Krisciunas et al. 2001),
which appear to be unreddened in their hosts, we can correct the
$V-J$, $V-H$, and $V-K$ colors for the effect of dust in our Galaxy (Schlegel
et al. 1998) and construct unreddened loci for slowly declining Type
Ia SNe.  We also used the optical data of SN 1999ee (Stritzinger et al.
2002) along with some unpublished IR data of SNe 1999ee and 2001ba
taken at Las Campanas Observatory and CTIO to constrain the 
shape of the unreddened loci.  

Following Krisciunas et al. (2000), let $t$ equal the number of days
since the time of $B$-band maximum, and T$_c$ be some ``crossover time'' when
the slope in the color curve has a sudden change.  We consider only the
data from $-9 < t < 27$ d and construct the simplest form of
unreddened loci that can fit the data.  For $V-J$ we have:

\begin {equation}
 V \; - \; J  \; = \; a_0 \; + \; a_i \; + \; 
b_1 \; (t - T_c) \; + \; c \; (t - T_c)^2 \; ; \; (t \; < \; T_c) \; .
\end {equation}

\begin {equation}
V \; - \; J  \; = \; a_0 \; + \; a_i \; + \; b_2 \; (t - T_c) \; ; \; (t \geq T_c) \; .
\end {equation}

\parindent = 0 mm

Each supernova has its own a$_i$ value, but the objects presumed to
be unreddened and which are used to establish the unreddened locus
will have $\langle$ a$_i \rangle$ = 0. Reddened objects have a$_i >
0$.  Under the assumption that the color curves of a group of
unreddened supernovae are validly parameterized by a uniform locus,
and that the reddened objects exhibit the same locus simply shifted
to the red, the goal is to use all the data for a given color index
to solve for the crossover time, the zeropoint, the slopes, the
second order coefficient, and the color excesses of each reddened
object.

\parindent = 9 mm

In Fig. 10 we show the unreddened $V-J$ color locus for slowly
declining Type Ia SNe.  Using the $B-V$ color excesses given by
Schlegel et al. (1998) and E($V-J$) = 2.223 E($B-V$) from a standard
model of Galactic dust (Rieke \& Lebofsky 1985), we have subtracted
$V-J$ color excesses of 0.089, 0.072, and 0.124 mag, respectively,
from the observed $V-J$ colors of SNe 1999aa, 1999aw, and 1999gp.  
We assume that these three objects were unreddened in their host
galaxies. For SN 1999ee we derive E($V-J$) = 0.658 $\pm$ 0.022.  
This has been subtracted from the SN 1999ee data shown in Fig. 10.
The $V -$ near IR color loci given by Krisciunas et al. (2000) were
not extrapolated beyond $t$ = 27 d, and we have truncated our $V-J$
locus at that epoch, though on the basis of SN 1999aw we might have
extended the right-hand line further.

The $V-J$ photometry of slowly declining Type Ia SNe gives a
crossover time T$_c$ = 14.43 $\pm$ 0.33 d.  We find $a_0$ = $-1.96$,
$b_1$ = $-$0.07480 $\pm$ 0.00747 mag d$^{-1}$, $c$ = $-$0.00015 $\pm$
0.00030 mag d$^{-2}$, and $b_2$ = 0.06867 $\pm$ 0.00500 mag d$^{-1}$.  
Whereas the $c$ term was necessary for the mid-range decliners
studied by Krisciunas et al. (2000), we found that it was not
statistically significantly different than zero for the slow
decliners. We list it here only for reasons of completeness.

Given the optical color excess E($B-V$) = 0.082 for SN 2000cx and
the interstellar extinction law quantified by Rieke \& Lebofsky
(1985), we estimate the following color excesses: E($V-J$) = 0.182;
E($V-H$) = 0.210; and E($V-K$) = 0.226. 

In Fig. 11 we show the dereddened $V-J$ colors of SN 2000cx, along
with the unreddened color locus described above (the solid line).  
We also show for reference (as a dashed line) the unreddened color
locus for mid-range decliners given by Krisciunas et al. (2000).  
The big dip observed in the $J$-band leads to a very blue $V-J$ color
for SN 2000cx at $t$ = 10 d.  However, the $V-J$ colors match the
unreddened locus for slow decliners overlapping the time of maximum
light.  We note that conversion of the $J$-band data to the filter
system of Persson et al. (1998) would make our data up to 0\fm07
fainter at maximum light, assuming that the IR spectral evolution of
SN 2000cx was similar to SN 1999ee (see Krisciunas et al. 2003, Fig.
6).  This would make the $V-J$ data shown on the left hand side of
Fig. 11 up to 0\fm07 bluer.

A similar analysis of the $V-H$ colors of SNe 1999aw, 1999gp, and
1999ee gives a crossover time T$_c$ = 5.71 $\pm$ 0.27 d, $a_0$ =
$-1.60$, $b_1$ = $-$0.06872 $\pm$ 0.00304 mag d$^{-1}$, $c \equiv$ 0,
and $b_2$ = 0.07502 $\pm$ 0.00262 mag d$^{-1}$.  In this case we have
subtracted E($V-H$) = 0.083 mag from the observed colors of SN 1999aw
and 0.143 mag from the data of SN 1999gp, to account for the effect
of dust in our Galaxy. We derived a total color excess of E($V-H$) =
0.795 $\pm$ 0.023 for the reddening of SN 1999ee.  The E($V-J$) and
E($V-H$) color excesses of SN 1999ee are consistent with one unique
value (rather than two disjoint values) of A$_V$ = 0.944 $\pm$ 0.061,
which is a good consistency check relating to the assumption that
there exist uniform color loci.

In Fig. 12 we show a corresponding plot for dereddened $V-H$ colors
of SN 2000cx, with the unreddened loci for slow decliners (solid
line) and mid-range decliners (dashed line) shown.  For $V-H$ there is
good agreement between the SN 2000cx data and the unreddened locus for slow
decliners from $t$ = 10 to 27 d.  Conversion of the
$H$-band data to the filter system of Persson et al. (1998) would
change the data from $10 < t < 27$ ~d by up to 0\fm06, making them
brighter, and the $V-H$ colors redder.  Once again, this assumes that
the IR spectral evolution of SN 2000cx was similar to that of SN 1999ee.

In Fig. 13 we show the $V-K$ data of SN~2000cx, corrected only for the
reddening due to our Galaxy, along with dereddened data of SNe 1999aa,
1999ee, 1999gp, and 2001ba.  The $V-K$ colors of SN~2000cx are quite
similar to these other slow decliners, both pre- and post-maximum.  From
$10 < t < 21$ d SN~2000cx is the bluest object in this color index.

\section{The Bolometric Behavior of SN~2000cx}

The wide wavelength coverage of the $UBVRIJHK$ broadband magnitudes
allows us to construct ultraviolet/optical/near-infrared ``uvoir''
bolometric light curves.  Only a few papers have been published trying
to estimate the bolometric light curves 
of Type Ia supernovae. (See Leibundgut \& Suntzeff 2003
for a summary.) The calculated luminosities are not
the true bolometric luminosities, but represent the fraction of the
gamma rays produced in the radioactive decays of the synthesized
nuclides that are thermalized in the expanding debris nebula. As shown
by Leibundgut \& Pinto (1992) and Leibundgut (2000), a significant 
fraction of the gamma rays leak out of the supernova debris going from 
10 percent at the time of $B_{max}$ to over 50 percent 40 days after maximum.

Suntzeff (1996) used ultraviolet spectra, optical $UBVRI$, and
near-infrared $JHK$ data to estimate accurate bolometric fluxes. Because
of the limited infrared data at the time, only a few points on the
bolometric light curves could be accurately calculated. Vacca \&
Leibundgut (1996) and Contardo, Leibundgut, \& Vacca (2000) obtained more
complete uvoir bolometric light curves by integrating optical broad-band
magnitudes.  Applying Arnett's law (Arnett 1982) to the peak bolometric
luminosities, Vacca \& Leibundgut (1996) and Contardo et al. (2000)  
found a range of more than a factor in 10 in the $^{56}$Ni masses for a
group of nearby Type Ia SNe.  Cappellaro et al. (1997) used a $V$
magnitude with a bolometric correction to study the gamma-ray trapping in
the late-time light curves, which also showed a significant range in
$^{56}$Ni masses.

We have integrated the broad-band magnitudes for SN~2000cx using a simple
trapezoidal rule and a conversion of broad-band to monochromatic fluxes
from Suntzeff \& Bouchet (1990). We have added extrapolations to the
ultraviolet from the $U$ filter and to the mid-infrared from the $K$ or
$H$ filter using an extrapolation scheme discussed by Suntzeff (2003).
These extrapolations only add $\sim2$ percent for the missing infrared
flux, and less than 10 percent for the ultraviolet after maximum light. In
Fig. 14 we plot the uvoir bolometric luminosity for SN~2000cx. For
comparison, we also plot similar data for SN~2001el and SN~1999ee taken
from Suntzeff (2003). (Note that the data have been offset slightly for
plotting purposes.) The SN~1999ee and SN~2001el data were taken from
Stritzinger et al. (2002) and Krisciunas et al. (2003) and are meant to
represent ``normal'' Type Ia supernovae with $\Delta$m$_{15}$($B$) of
0.94
and 1.13, which are similar to the $\Delta$m$_{15}$ value of SN~2000cx. We
have assumed the following distance moduli and $B-V$ reddening, which are
based on a distance scale of $H_0=74$ km s$^{-1}$ Mpc$^{-1}$: SN~2000cx,
(32.47,0.082); SN~1999ee, (33.21,0.30); SN~2001el, (31.26,
0.21).\footnote[4]{A systematic error of $\Delta$m in the distance modulus
corresponds to a change in $log$(L) of 0.4 $\times \Delta$m.  If the true
distance modulus of SN~2000cx is 31.90, then we must shift the bolometric
light curve of SN~2000cx in Figs. 14 and 15 down by 0.228 mag.} The
bolometric light curves have been plotted relative to the time of peak
bolometric luminosity (not $B_{max}$) which we estimated from the
curves.

% ccc

The only remarkable difference between SN~2000cx and the normal Type
Ia's as seen in Fig. 14 is the lack of the flux excess around
day 30. This flux excess, which is associated with the secondary
maxima in $IJHK$ and a ``shoulder'' in $R$ was noted by
Suntzeff (1996) and Contardo et al. (2000). Since the energy source for
the bolometric
luminosity and the optical depth to gamma rays are monotonically
declining at that date, such an inflection in the bolometric
luminosity must be associated with cooling mechanism and not the
energy input to the nebula. 

In Fig. 15 we can see the differences in the bolometric light
curves more clearly. Here we plot smoothed representations of the
bolometric light curves, and at the bottom of the panel, the
bolometric light curve of SN~2000cx with respect to SN~1999ee and
SN~2001el. It can be seen that SN~2000cx rises to maximum more quickly
and falls more quickly (within 10 days of maximum), but the effect is
rather small. However, starting around day 25, SN~2000cx suddenly
declines rapidly compared to these two SNe, reaches a maximum flux
deficit around day 35, and then increases in brightness slightly to
day 50. Fig. 15 gives the impression that the post-maximum
flux enhancement of SN~2000cx was weaker and earlier than the
comparison Type Ia events.

Two explanations for the secondary maximum have been published. One
explanation for the secondary $IJHK$ maxima has been given by
Pinto \& Eastman (2000a, 2000b) who note that this flux enhancement is due
to a rapid change in the flux mean opacity. After maximum light, the
thermalized energy input to the light curve from the radioactive
nuclides is less than the observed luminosity, implying qualitatively
that the post-maximum luminosity is powered by a reservoir of
previously trapped radiation. If the post-maximum opacity decreases
due to a drop in the effective temperature, the diffusion times drop
and the trapped energy escapes more rapidly leading to a pause in the
rapid luminosity decline. This bolometric flux excess appears in the
redder colors because the opacities are very low and there are ample
emission sources such as \ion{Fe}{2} and \ion{Ca}{2}.

A similar explanation has been given by H\"{o}flich (1995) and
H\"{o}flich, Khokhlov, \& Wheeler (1995). 
They note that the infrared luminosity can be roughly approximated 
by the Rayleigh-Jeans limit, namely:

\begin{equation}
(L_{IR-color}) \propto R^2_{ph}T_{eff}D_{IR}
\end{equation}

\noindent Here $R_{ph}$, $T_{eff}$, and $D_{IR}$ are the photospheric
radius, the effective temperature, and a dilution factor appropriate
for a scattering dominated atmosphere. They argue that $D_{IR}$ is a
slowly varying function during this epoch. They show that after
maximum, $T_{eff}$ drops as the energy source switches from $^{56}$Ni
to $^{56}$Co which starts the steep post-maximum decline. In many of
their models however, the photospheric radius $R_{ph}$ increases as it
is dragged out by the expansion of the debris. Depending on the rate
of expansion of the photosphere, this can cause the product in the
equation for $L_{IR-color}$ to increase. 
The appearance of an expanding photosphere can only be maintained if
the opacities stay high. In these conditions, as the photosphere is
pulled outward with the debris, the luminosity will increase if
$T_{eff}$ does not drop dramatically. This high rate of cooling, which
greatly exceeds the energy input from the radioactive nuclides, cannot
be maintained indefinitely, and at some point the opacities will begin
to drop so rapidly that apparent photospheric radius will also being
to recede quickly causing a sudden decreace in luminosity.

According to the models discussed above, this would support the
hypothesis that SN~2000cx is a sub-luminous event. The sub-luminous
SNe 1992A, 1992bo, and 1991bg
(Contardo et al. 2000; see their Figure 5) also have weak or no
secondary flux enhancements and SN~2000cx clearly belongs to this
class.

Unfortunately, given the large number of unknown parameters in the models
for the explosions of Type Ia SNe, this is not a strong conclusion. Pinto
\& Eastman (2000b) note that the secondary maximum is a sensitive function
of how the much of the radioactive nuclides are mixed in the expanding
debris from the core. An unusual mixing event bringing relatively more
$^{56}$Ni out from the core qualitatively could account for the lack of
the secondary flux enhancement independent of the intrinsic luminosity.

Returning to the empirical bolometric light curve, we find two other
aspects of the morphology of the bolometric light curve which point to
this event being sub-luminous. Contardo et al. (2000) showed that the
decline rate between days 50 and 80 for their sample of Type Ia supernovae
was $2.6\pm0.1$ mag per 100 days, except for most sub-luminous event in
their sample, SN~1991bg, which had a decline rate of 3.0 mag per 100 days.
SN~2000cx declined at 2.9 mag per 100 days during this time period. If one
looks at their Fig. 5, one can also see that SN~1991bg also stands out in
the peak-to-tail luminosity difference. In Fig. 16 we plot the difference
in luminosity between the peak luminosity and 90 days after peak for the
sample studied by Contardo et al. (2000). Evidently, this luminosity
difference is correlated with $\Delta$m$_{15}$($B$), with the fainter SNe
having larger luminosity differences. The observed value of this
luminosity difference in SN~2000cx of 1.63 associates it with the
sub-luminous group. This is also seen in Fig. 15 where this SN is compared
to SNe 2001el and 1999ee. In that figure, the luminosity on the
exponential tail for days beyond 50 are underluminous by about 0.2dex with
respect to SN~2001el, which has a value of $\Delta$m$_{15}$($B$) of 1.13.

By day 50 or so, the energy deposition in a typical Type Ia supernova
due to the thermalization of gamma rays occurs in regions which are
optically thin in the optical and near-infrared
(Pinto \& Eastman 2000a). Thus, the luminosity at this epoch responds
rapidly to the input energy source, which at this time is
$^{56}$Co. At maximum light, according to Arnett's law, the bolometric
luminosity equals the instantaneous energy input from the radioactive
nuclides.  The ratio of these two luminosities should then be
independent, to first order, of the mass of $^{56}$Ni synthesized.

The larger luminosity difference between the peak and 90 days after peak
is thus caused by a smaller optical depth to gamma
rays at late times. This could be caused, for instance, by positron
escape as discussed by Milne, The, \& Leising (2001) but the modelling
shows minimal effects of positron trapping at this epoch due to the short
positron lifetimes which approximate in situ deposition of positron
energy. A lower optical depth to gamma rays could also be caused by an
asymmetric mass distribution of the ejecta.  Perhaps the simplest way to
lower the optical depth is to increase the kinetic energy due to the
explosion.

It is not unreasonable that the initial kinetic energy may be only
vaguely related to the mass of $^{56}$Ni synthesized. 
Pinto \& Eastman (2000a; see their Figure 4), showed that a {\it larger}
kinetic energy
will lead to a more rapid decline in the mass column depth, and
produce a narrower bolometric light curve and increased the
peak-to-tail luminosity difference. In a subsequent article,
Pinto \& Eastman (2001) note that the explosion kinetic energy is not
necessarily a function of the $^{56}$Ni mass, since the total energy of
the burning of a C/O mixture is nearly the same if it burns to the Si
group or to $^{56}$Ni. The thermal energy liberated at the explosion
does go entirely to the kinetic energy, but that kinetic energy may
not be strongly coupled to the $^{56}$Ni, which powers the subsequent
light curve. Li et al. (2001) found that SN~2000cx did have very high
sulfur and silicon velocities compared to SN~1994D, and also concluded
that SN~2000cx may have had a larger than typical kinetic energy.

Thus, if the SBF distance is correct, a working hypothesis to explain
the bolometric behavior of SN~2000cx is that this supernova is an
{\em underluminous event with higher than normal kinetic energy}. We should
not be too forceful in stressing this conclusion however. This event
was clearly unusual in its color evolution, and simple morphological
arguments may fail if model parameters, such as the extent of mixing
of the radioactive nuclides or the symmetry of the explosion, are
uncoupled from the other fundamental parameters such as the amount of
$^{56}$Ni synthesized. None of the evidence here clearly points to a
shorter or longer distance to this supernova. Obviously, we need a
better distance to NGC~524 to resolve this question.

% bbb

\section{Conclusions}

SN 2000cx, the brightest supernova discovered in the year 2000, occurred 
in the unobscured outer regions of an early-type galaxy and was well observed with
multiple telescopes, allowing us to compile a datset of unprecedented
size.  While many Type Ia SNe have light curves that follow patterns
that are now well established, SN 2000cx did not conform to these
patterns.  

SN 2000cx was a reasonably fast riser in $B$ and $V$.  From the pre-maximum photometry
Li et al. (2001) obtain a stretch factor that corresponds to $\Delta$m$_{15}$($B$)
= 1.64 $\pm$ 0.02.  Based solely on its weak
$I$-band secondary hump, we would have predicted $\Delta$m$_{15}$($B$) $\approx$
1.7.  If the distance modulus based on Surface Brightness Fluctuations of
the host galaxy is correct (Tonry et al. 2001), the corresponding absolute 
magnitudes in $BVIH$ are comparable to fast decliners, with $\Delta$m$_{15}$($B$)
in the range 1.4 to 1.7.

However, SN 2000cx was a slow {\em decliner}, with $\Delta$m$_{15}$($B$) = 0.93.
Its pre-maximum spectrum showed strong Fe III and weak Si II, like other
slow decliners such as SN 1991T.  Its $V-K$ color evolution, both pre- and
post-maximum, was very similar to that of other slow decliners.

The bolometric behavior of SN~2000cx, when compared to the normal SNe
1999ee and 2001el, showed that this SN rose and fell from maximum
light more rapidly, and that the magnitude difference between peak
brightness and 90 days past peak was larger than normal. This behavior
is consistent with the higher kinetic energies seen at maximum light
and reported by Li et al. (2002), but it can also be explained by this
event being sub-luminous.

The distance modulus of SN 2000cx is somewhat problematic. We note,
however, that MLCS (Riess et al. 1996, 1998), Hubble Law's (with H$_0$ = 74
km s$^{-1}$ Mpc$^{-1}$), and the method that uses the $H$-band absolute
magnitude at $t$ = 10 d (Krisciunas et al. 2003) give just about the same
value.  The SBF method gives a distance modulus roughly 0.6 mag smaller.  

Given: 1) the accuracy of the photometry of SN~2000cx at maximum ($\pm$ 0.03 mag);
2) the host extinction of SN~2000cx
is minimal (or zero); 3) the uncertainty of the Galactic extinction correction
is also small; and 4) the light curves cannot be fit by templates based on 
other objects, confidently placing this SN in a Hubble diagram depends significantly
on a direct measure of the host galaxy's distance (such as with the SBF
method) or on an absolute magnitude derived from an explosion model that 
can match the many unusual observed facts.

\acknowledgements  

Support for Proposal Number GO-07505.02A, GO-08177.6, and
GO-08641.07A was provided by NASA through a grant from the Space
Telescope Science Institute, which is operated by the Association of
Universities for Research in Astronomy, Inc., under NASA contract
NAS5-26555. This paper is based, in part, on observations obtained
with the Apache Point Obsevatory 3.5m telescope, which is owned and
operated by the Astrophysical Research Consortium.

We thank W. D. Li for sharing the KAIT data considerably ahead of publication.
We thank M. R. Garcia, L. Clark, D. Hoard, J. Alfonso, and K. Vivas for
obtaining some of the data at CTIO.  C. Stubbs, G. Miknaitis, and E. Bergeron
helped with data acquisition for some APO observations.  D. Edgeworth helped
with some of the MRO observations. The APO infrared data were reduced
using software written in part by A. Diercks and E. Magnier.

\appendix

\section{The $R$-band filter used at YALO}

In Fig. 17 we show the filter transmission function of the $R$-band filter
used with the YALO 1-m telescope for the observations presented here.  
This is a much more standard filter than that used by Stritzinger et al.
(2002) for their observations of SN~1999ee using the same telescope and
camera.  Also shown in the diagram is the {\em effective} throughput in
$R$, made up of the filter transmission function multiplied by an
atmospheric transmission function, the quantum efficiency of the chip, and
two aluminum reflections.  Both functions shown in Fig. 17 include a 280
\AA\ shift to the blue, which comprises the 310 \AA\ shift suggested by
the manufacturer, taking into account the use of the filter when cooled,
and a 30 \AA\ shift back to red.  This smaller shift was necessary to
match the color term obtained for $R$-band photometry based on synthetic
photometry with the actual color term derived from the observations of
Landolt (1992) standards.

\clearpage

\begin{deluxetable}{crcccc}
\tablecolumns{6}
\tablewidth{0pc} 
\tablecaption{Average Photometric Transformation Values$^a$}
\startdata 
Parameter   &  Color Term &    Mean Error  &  Color Index  &  Extinction  & Mean Error \\ \hline
\multicolumn{6}{c}{YALO 1-m (3 nights)} \\ \hline
$B$           &  $-$0.079   &  0.003         &   $B-V$         &  0.248       & 0.007 \\
$V$           &  0.018      &  0.003         &   $B-V$         &  0.131       & 0.005 \\
$R$           &  $-$0.030   &  0.005         &   $V-R$         &  0.076       & 0.007 \\
$I$           &  0.045      &  0.003         &   $V-I$         &  0.009       & 0.006 \\ \hline
\multicolumn{6}{c}{CTIO 0.9-m (7 nights)} \\ \hline
$U$    & $-$0.076  & 0.016   & $U-V$   &  0.443       &  0.054 \\
$B$    & 0.126	   & 0.009   & $B-V$   &  0.236       &  0.018 \\
$V$    & $-$0.017  & 0.006   & $B-V$   &  0.139       &  0.009 \\
$R$    & 0.005	   & 0.014   & $V-R$   &  0.087       &  0.011 \\
$I$    & $-$0.007  & 0.009   & $V-I$   &  0.048       &  0.012 \\ \hline
\multicolumn{6}{c}{MRO (15 nights)} \\ \hline
$V$     & 0.044   & 0.003   &    $b-v$   & 0.201    & 0.029 \\
$B-V$   & 1.067   & 0.004   &    $b-v$   & 0.142    & 0.015 \\
$V-R$   & 1.026   & 0.006   &    $v-r$   & 0.061    & 0.021 \\
$V-I$   & 1.014   & 0.003   &    $v-i$   & 0.083    & 0.027 \\ \hline
\multicolumn{6}{c}{APO (9 nights)} \\ \hline
$V$     & 0.026   & 0.011    &  $b-v$   &  0.216   &  0.029 \\
$B-V$   & 0.975   & 0.002    &  $b-v$   &  0.103   &  0.016 \\
$V-R$   & 1.040   & 0.003    &  $v-r$   &  0.036   &  0.024 \\
$V-I$   & 1.003   & 0.017    &  $v-i$   &  0.083   &  0.030 \\
\enddata
\tablenotetext{a} {The color terms for YALO and the CTIO 0.9-m scale standardized
colors, while the color terms for APO and MRO scale instrumental colors.} 
\end{deluxetable}

\clearpage

\begin{deluxetable}{cccccc}
\tablecolumns{6}
\tablewidth{0pc} 
\tablecaption{Photometry of Comparison Stars}
\startdata
ID & $V$             & $B-V$        &  $U-B$       & $V-R$        &  $V-I$  \\ \hline \hline
1$^a$  & 11.141 (0.001) & 0.536 (0.002) & 0.000 (0.004) & 0.328 (0.004) & 0.666 (0.002)  \\
1$^b$      & 11.145 (0.005) & 0.528 (0.011) & \ldots        & 0.325 (0.011) & 0.672 (0.010) \\
2$^a$  & 12.564 (0.002) & 0.606 (0.011) & 0.015 (0.021) & 0.366 (0.002) & 0.733 (0.004)       \\
2$^b$      & 12.572 (0.004) & 0.587 (0.006) &   \ldots      & 0.370 (0.009) & 0.749 (0.007)   \\
3$^a$  & 14.780 (0.005) & 0.778 (0.012) & 0.172 (0.038) & 0.441 (0.007) & \ldots             \\
4$^a$  & 13.494 (0.003) & 0.797 (0.004) & 0.315 (0.006) & 0.470 (0.004) & 0.911 (0.005)       \\ 
4$^b$  & 13.509 (0.003) & 0.778 (0.008) &  \ldots       & 0.471 (0.007) & 0.930 (0.007)  \\
5$^a$  & 15.364 (0.007) & 1.005 (0.017) & 0.671 (0.050) & 0.545 (0.009) & \ldots             \\  
6$^a$  & 16.366 (0.005) & 1.044 (0.017) & 0.903 (0.072) & 0.624 (0.007) & 1.167 (0.007)       \\
7$^a$  & 16.765 (0.007) & 0.962 (0.014) & 0.712 (0.176) & 0.584 (0.009) & 1.096 (0.009)       \\
8$^a$  & 12.812 (0.006) & 0.696 (0.011) & 0.099 (0.022) & 0.400 (0.007) & 0.797 (0.008)       \\
8$^b$  & 12.808 (0.005) & 0.677 (0.008) &  \ldots       & 0.384 (0.024) & 0.800 (0.006)  \\
9$^a$  & 15.223 (0.003) & 1.233 (0.013) & 1.211 (0.041) & 0.727 (0.005) & 1.339 (0.005)       \\
10$^a$ & 14.559 (0.003) & 0.826 (0.009) & 0.352 (0.022) & 0.463 (0.005) & 0.908 (0.004)       \\
11$^a$ & 15.616 (0.008) & 1.158 (0.022) & 1.233 (0.086) & 0.687 (0.010) & \ldots             \\ 
\hline
        &    $J$          &     $H$       &     $K$       &   &  \\
1$^c$   &  10.006 (0.006) & 9.758 (0.008) & 9.689 (0.008) &  &  \\
2$^c$   &  11.974 (0.006) & 11.585 (0.015) & 11.501 (0.014) & & \\
\enddata
\tablenotetext{a} {Mean values from 7 nights of CTIO 0.9-m imagery.
$^b$Mean values from 6 nights of MRO imagery.
$^c$Mean values from 5 nights of LCO imagery.}
\end{deluxetable}

\clearpage

\begin{deluxetable}{ccccccc}
\tablecolumns{7}
\rotate
\tablewidth{0pc} 
\tablecaption{$UBVRI$ Photometry of SN 2000cx. CTIO 0.9-m Data$^a$}
\startdata 
JD+2451000 & $U$            &      $B$         &    $V$            &     $R$           &     $I$             & Observer  \\ \hline \hline
745.88  &   13.722 (0.011)  &  13.995 (0.007)  &  13.821 (0.004)  &  13.796 (0.006)  &  \ldots         & Leiton/Clark \\ 
749.84  &   13.223 (0.015)  &  13.517 (0.012)  &  13.380 (0.008)  &  13.443 (0.013)  &  13.617 (0.014) & Garcia \\  
750.86  &   13.195 (0.011)  &  13.474 (0.008)  &  13.336 (0.005)  &  13.408 (0.056)  &  \ldots         & Garcia \\  
751.85  &   13.190 (0.011)  &  13.442 (0.008)  &  13.294 (0.005)  &  13.380 (0.010)  &  13.672 (0.011) & Garcia \\  
757.84  &   \ldots          &  13.640 (0.008)  &  13.328 (0.005)  &  13.444 (0.007)  &  13.949 (0.011) & Hoard \\  
824.70  &   \ldots          &  17.134 (0.041)  &  16.672 (0.018)  &  16.583 (0.028)  &  17.018 (0.039) & Smith \\ 
825.71  &   17.342 (0.064)  &  17.157 (0.033)  &  16.687 (0.020)  &  16.622 (0.032)  &  17.103 (0.044) & Smith \\
826.67  &   17.394 (0.094)  &  17.226 (0.042)  &  16.720 (0.023)  &  16.651 (0.042)  &  17.140 (0.056) & Smith \\
827.72  &   \ldots          &  17.225 (0.009)  &  16.752 (0.004)  &    \ldots        &  17.175 (0.019) & Smith \\ 
834.65  &   \ldots          &  17.356 (0.026)  &  16.938 (0.017)  &    \ldots        &  17.490 (0.029) & Candia \\
\enddata
\tablenotetext{a} {The uncertainties given are due to photon statistics and
zeropoint errors only.}
\end{deluxetable}

\clearpage

\begin{deluxetable}{cccccc}
\tablecolumns{6}
\rotate
\tablewidth{0pc} 
\tablecaption{$BVRI$ Photometry of SN 2000cx. YALO Data$^a$}
\startdata 
JD+2451000 &          $B$      &      $V$         &     $R$          &     $I$              & Observer \\ \hline \hline
744.90     &  14.158 (0.083)  & 14.116 (0.059)  & 13.973 (0.095)  &  14.037 (0.084)      & D. Gonzalez \\  
745.86     &  13.955 (0.008)  & 13.790 (0.005)  & 13.774 (0.016)  &  13.875 (0.011)      & D. Gonzalez \\   
750.88     &  13.419 (0.010)  & 13.288 (0.007)  & 13.369 (0.014)  &  13.643 (0.012)      & D. Gonzalez \\   
752.89     &  13.409 (0.015)  & 13.230 (0.006)  & 13.361 (0.032)  &  13.820 (0.076)      & J. Espinoza \\  
755.87     &  13.496 (0.009)  & 13.240 (0.007)  & 13.359 (0.019)  &  13.873 (0.014)      & J. Espinoza \\   
757.82     &  13.602 (0.018)  & 13.296 (0.010)  & 13.439 (0.020)  &  13.960 (0.016)      & J. Espinoza  \\  
760.81     &  13.810 (0.066)  & 13.440 (0.017)  & 13.673 (0.029)  &  14.309 (0.040)      & J. Espinoza \\   
763.84     &  14.033 (0.007)  & 13.639 (0.003)  & \ldots         &   14.647 (0.020)      & D. Gonzalez  \\  
766.83     &  14.305 (0.007)  & 13.862 (0.005)  & 14.165 (0.013)  &  14.798 (0.011)      & J. Espinoza \\   
769.84     &  14.602 (0.012)  & 14.061 (0.008)  & 14.265 (0.053)  &  14.777 (0.016)      & J. Espinoza \\   
772.74     &  14.915 (0.013)  & 14.234 (0.008)  & 14.301 (0.024)  &  14.656 (0.016)      & D. Gonzalez \\   
775.77     &  15.263 (0.023)  & 14.444 (0.016)  & 14.355 (0.046)  &  14.522 (0.036)      & D. Gonzalez \\   
781.77     &  15.958 (0.006)  & 15.001 (0.003)  & 14.758 (0.020)  &  14.640 (0.016)      & J. Espinoza \\   
784.77     &  16.191 (0.009)  & 15.232 (0.003)  & 14.970 (0.016)  &  14.881 (0.009)      & J. Espinoza \\   
791.75     &  16.491 (0.008)  & 15.583 (0.003)  & 15.332 (0.006)  &  15.355 (0.008)      & D. Gonzalez \\   
802.74     &  16.741 (0.028)  & 16.002 (0.018)  & 15.757 (0.039)  &  15.928 (0.030)      & D. Gonzalez \\   
805.83     &  16.777 (0.014)  & 16.084 (0.005)  & 15.873 (0.011)  &  16.081 (0.013)      & D. Gonzalez \\   
812.85     &  16.912 (0.007)  & 16.293 (0.004)  & 16.115 (0.010)  &  16.401 (0.010)      & J. Espinoza \\   
820.70     &  17.032 (0.018)  & 16.539 (0.010)  & 16.394 (0.017)  &  16.784 (0.019)      & J. Alfonso \\  
827.68     &  17.154 (0.020)  & 16.760 (0.009)  & 16.613 (0.014)  &  17.045 (0.016)      & J. Espinoza/K. Vivas  \\   
\enddata
\tablenotetext{a} {The uncertainties given are due to photon statistics and 
zeropoint errors only.
From polynomial fits to the YALO data we estimate that the internal
errors are greater than or equal to $\sigma _B = \pm$ 14; $\sigma _V = \pm$ 34; $\sigma _R =
\pm$ 41; and $\sigma_I = \pm$ 71 mmag. See text for a discussion of systematic errors.}
\end{deluxetable}

\clearpage

\begin{deluxetable}{cccccc}
\tablecolumns{6}
\rotate
\tablewidth{0pc} 
\tablecaption{$BVRI$ Photometry of SN 2000cx. APO Data$^a$}
\startdata 
JD+2451000 &          $B$     &      $V$          &     $R$           &     $I$          &   Observers \\ \hline \hline
760.90    &   13.829 (0.005)  &   13.512 (0.002)  &   13.757 (0.003)  &   14.270 (0.005) &   Krisciunas/McMillan       \\     
784.70    &   16.169 (0.014)  &   15.263 (0.004)  &   15.012 (0.010)  &   14.799 (0.010) &   Rest/McMillan             \\   
788.71    &   16.336 (0.008)  &   15.473 (0.003)  &   15.214 (0.008)  &   15.081 (0.008) &   Rest/Miknaitis/Hastings   \\  
822.75    &   17.066 (0.010)  &   16.635 (0.004)  &   16.515 (0.009)  &   16.861 (0.013) &   Krisciunas/McMillan       \\ 
834.61    &   \ldots          &   16.976 (0.007)  &   \ldots          &   17.299 (0.017) &   Krisciunas/Bergeron       \\ 
\enddata
\tablenotetext{a} {The uncertainties given are due to photon statistics and 
zeropoint errors only.}
\end{deluxetable}

\begin{deluxetable}{cccccc}
\tablecolumns{6}
\rotate
\tablewidth{0pc} 
\tablecaption{$BVRI$ Photometry of SN 2000cx. MRO Data$^a$}
\startdata 
JD+2451000 &          $B$     &      $V$          &     $R$           &     $I$           &   Observer \\ \hline \hline
744.92    &   14.177 (0.017)  &   14.008 (0.007)  &   13.997 (0.011)  &   14.057 (0.015)  &   Tavenner/Logan      \\           
745.93    &   13.975 (0.012)  &   13.814 (0.005)  &   13.788 (0.008)  &   13.869 (0.011)  &   Logan               \\                    
752.88    &   13.444 (0.006)  &   13.281 (0.002)  &   13.409 (0.004)  &   13.719 (0.007)  &   Krisciunas          \\               
753.96    &   13.458 (0.007)  &   13.263 (0.003)  &   13.389 (0.004)  &   13.729 (0.009)  &   Thomas/Snider       \\             
754.97    &   13.487 (0.005)  &   13.265 (0.002)  &   13.384 (0.004)  &   13.757 (0.007)  &   Thomas/Snider       \\             
757.94    &   13.640 (0.006)  &   13.346 (0.002)  &   13.498 (0.004)  &   13.968 (0.006)  &   Tavenner/Logan      \\           
758.94    &   13.695 (0.007)  &   13.380 (0.003)  &   13.558 (0.005)  &   14.039 (0.008)  &   Tavenner/Logan      \\           
759.93    &   13.768 (0.006)  &   13.423 (0.003)  &   \ldots          &   14.149 (0.008)  &   Krisciunas/Edgeworth  \\     
764.91    &   14.162 (0.007)  &   13.803 (0.003)  &   14.135 (0.005)  &   14.591 (0.009)  &   Tavenner/Logan      \\           
765.96    &   14.250 (0.005)  &   13.887 (0.002)  &   14.219 (0.004)  &   14.636 (0.007)  &   Tavenner/Logan      \\           
769.88    &   14.626 (0.007)  &   14.142 (0.003)  &   14.412 (0.006)  &   14.619 (0.008)  &   Krisciunas          \\             
776.91    &   15.424 (0.021)  &   14.639 (0.011)  &   14.516 (0.020)  &   14.494 (0.025)  &   G. Gonzalez/West       \\           
783.00    &   16.094 (0.028)  &   15.140 (0.012)  &   14.820 (0.023)  &   14.728 (0.034)  &   Tavenner/Logan      \\          
784.95    &   16.166 (0.015)  &   15.284 (0.006)  &   15.033 (0.013)  &   14.788 (0.015)  &   Tavenner/Logan      \\          
807.83    &   16.860 (0.017)  &   16.170 (0.007)  &   16.097 (0.020)  &   16.209 (0.018)  &   Krisciunas          \\            
\enddata
\tablenotetext{a} {The uncertainties given are due to photon statistics and 
zeropoint errors only.
From polynomial fits to the MRO data we estimate that the internal
errors are greater than or equal to $\sigma _B = \pm$ 6; $\sigma _V = \pm$ 20; $\sigma _R =
\pm$ 43; and $\sigma_I = \pm$ 51 mmag. See text for a discussion of systematic errors.}
\end{deluxetable}

\clearpage

\begin{deluxetable}{ccccc}
\tablecolumns{5}
\tablewidth{0pc} 
\tablecaption{$JHK$ Photometry of SN 2000cx. YALO Data}
\startdata 
 JD+2451000 &       $J$         &         $H$       &      $K$           &    Observer  \\ \hline  \hline
   744.89  &   14.280 (0.009)  &   14.378 (0.017)  &   14.324 (0.031)  &   D. Gonzalez\\
   745.86  &   14.118 (0.008)  &   14.123 (0.008)  &   14.254 (0.024)  &   D. Gonzalez\\
   750.87  &   13.900 (0.007)  &   14.071 (0.014)  &   13.949 (0.022)  &   D. Gonzalez\\
   752.89  &   14.052 (0.008)  &   14.241 (0.016)  &   14.112 (0.025)  &   J. Espinoza\\
   755.86  &   14.325 (0.011)  &   14.373 (0.023)  &   14.195 (0.028)  &   J. Espinoza\\
   757.82  &   14.569 (0.011)  &   14.385 (0.019)  &   14.377 (0.031)  &   J. Espinoza\\
   760.81  &   15.574 (0.058)  &   14.659 (0.054)  &   \ldots          &   D. Gonzalez\\
   763.84  &   15.879 (0.025)  &   14.693 (0.026)  &   14.585 (0.037)  &   D. Gonzalez\\ 
   766.83  &   16.034 (0.023)  &   14.702 (0.029)  &   14.679 (0.032)  &   J. Espinoza\\
   769.84  &   16.085 (0.052)  &   \ldots          &   14.546 (0.045)  &   J. Espinoza\\
   772.74  &   15.861 (0.028)  &   14.550 (0.020)  &   14.638 (0.038)  &   D. Gonzalez\\    
   775.77  &   15.486 (0.015)  &   14.425 (0.016)  &   14.402 (0.032)  &   D. Gonzalez\\  
   781.78  &   15.470 (0.018)  &   14.660 (0.017)  &   14.673 (0.029)  &   J. Espinoza\\
   784.77  &   15.712 (0.018)  &   14.896 (0.020)  &   14.997 (0.040)  &   J. Espinoza\\
   787.76  &   16.028 (0.037)  &   15.136 (0.044)  &   15.034 (0.083)  &   D. Gonzalez\\
   791.74  &   16.376 (0.029)  &   15.261 (0.027)  &   15.662 (0.119)  &   D. Gonzalez\\
   802.73  &   17.408 (0.094)  &     \ldots        &   \ldots          &   D. Gonzalez\\
   812.84  &     \ldots        &   16.519 (0.091)  &   \ldots          &   J. Espinoza\\
\enddata
\end{deluxetable}

\clearpage

\begin{deluxetable}{ccccc}
\tablecolumns{5}
\tablewidth{0pc} 
\tablecaption{ $JHK$ Photometry of SN 2000cx. APO  Data}
\startdata 
JD+2451000 &   $J$          &    $H$          &       $K$         &   Observers  \\ \hline \hline
760.94  &   \ldots          & 14.484 (0.025)  &   14.410 (0.110)  & Krisciunas/McMillan \\  
771.78  &   15.736 (0.047)  & 14.541 (0.023)  &   14.398 (0.033)  & Krisciunas/Hastings/McMillan \\   
777.88  &   15.153 (0.054)  & 14.452 (0.022)  &   14.606 (0.088)  & Stubbs/McMillan/Hastings  \\   
778.76  &   15.136 (0.025)  & 14.483 (0.023)  &   14.500 (0.048)  & Krisciunas/McMillan/Hastings  \\   
784.73  &   15.533 (0.037)  & 14.947 (0.027)  &   14.867 (0.064)  & Rest/McMillan  \\   
788.77  &   15.909 (0.024)  & 15.287 (0.022)  &   15.326 (0.036)  & Rest/Miknaitis/Hastings  \\   
834.76  &   18.149 (0.134)  & 17.910 (0.243)  &   \ldots          & Krisciunas/Bergeron  \\    
\enddata
\end{deluxetable}

\begin{deluxetable}{cccc}
\tablecolumns{5}
\tablewidth{0pc} 
\tablecaption{Corrections from APO to YALO for $JHK$ bands$^a$ }
\startdata 
JD+2451000 &   $\Delta J$  &   $\Delta H$  &  $\Delta K$  \\ \hline \hline
760.94     & \ldots      & 0.040        &   0.005 \\
771.78     & 0.123       & 0.056        &   0.015  \\   
777.88     & 0.096       & 0.056        &   0.000 \\   
778.76     & 0.088       & 0.053        &   $-$0.001 \\   
784.73     & 0.117       & 0.011        &   0.002 \\   
788.77     & 0.153       & $-$0.021       &   0.004  \\   
\enddata
\tablenotetext{a} {These values are to be {\em added} to the APO data.}
\end{deluxetable}

\begin{deluxetable}{ccc}
\tablecolumns{3}
\tablewidth{0pc} 
\tablecaption{Maximum Magnitudes of SN 2000cx}
\startdata 
Filter &  JD+2451000    &   $ m_{max}$   \\ \hline \hline
$U$      &  751.9 (1.5)    &  13.19 (0.02)    \\
$B$      &  752.5 (0.5)    &  13.42 (0.02)    \\ 
$V$      &  753.7 (1.1)    &  13.25 (0.03)    \\
$R$      &  753.5 (1.1)    &  13.34 (0.04)    \\
$I$      &  750.2 (1.0)    &  13.65 (0.04)    \\
$J$      &  749.9 (0.3)    &  13.85 (0.10)     \\
$H$      &  749.0 (0.7)    &  14.04 (0.06)     \\
$K$      &  749.5 (0.2)    &  13.98 (0.04)     \\
\enddata
\end{deluxetable}

\clearpage

\begin{figure}
\figurenum{1}
\plotone{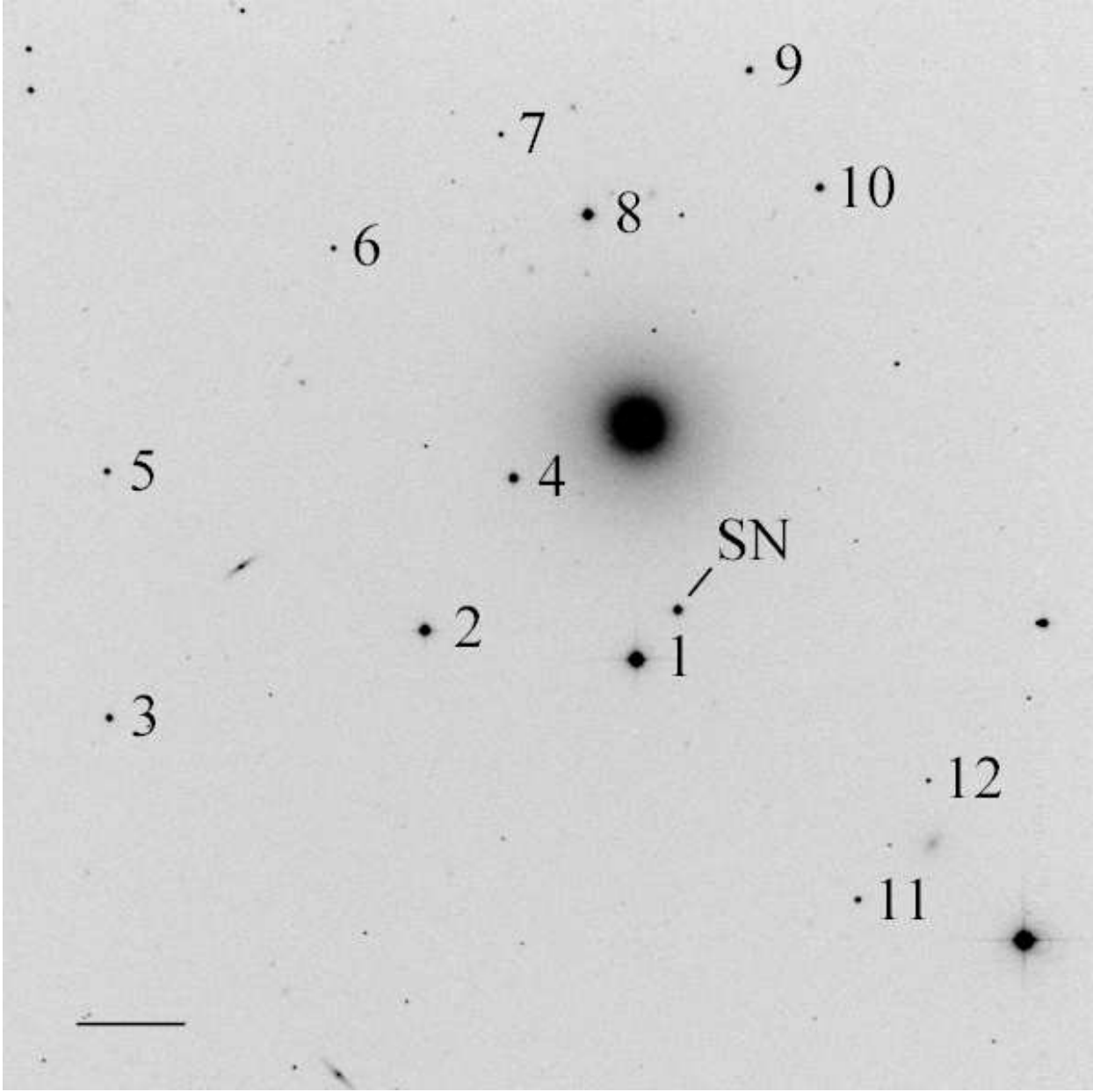}
\caption{Finding Chart for local standards near SN~2000cx in NGC 524. 
The image is a combined $BVR$ image taken with the CTIO 0.9-m
telescope. N is up and E is to the left. The horizontal bar shows a
scale of 1 arcminute.}
\end{figure}

\clearpage

\begin{figure}
\figurenum{2}
\epsscale{1}
\plotone{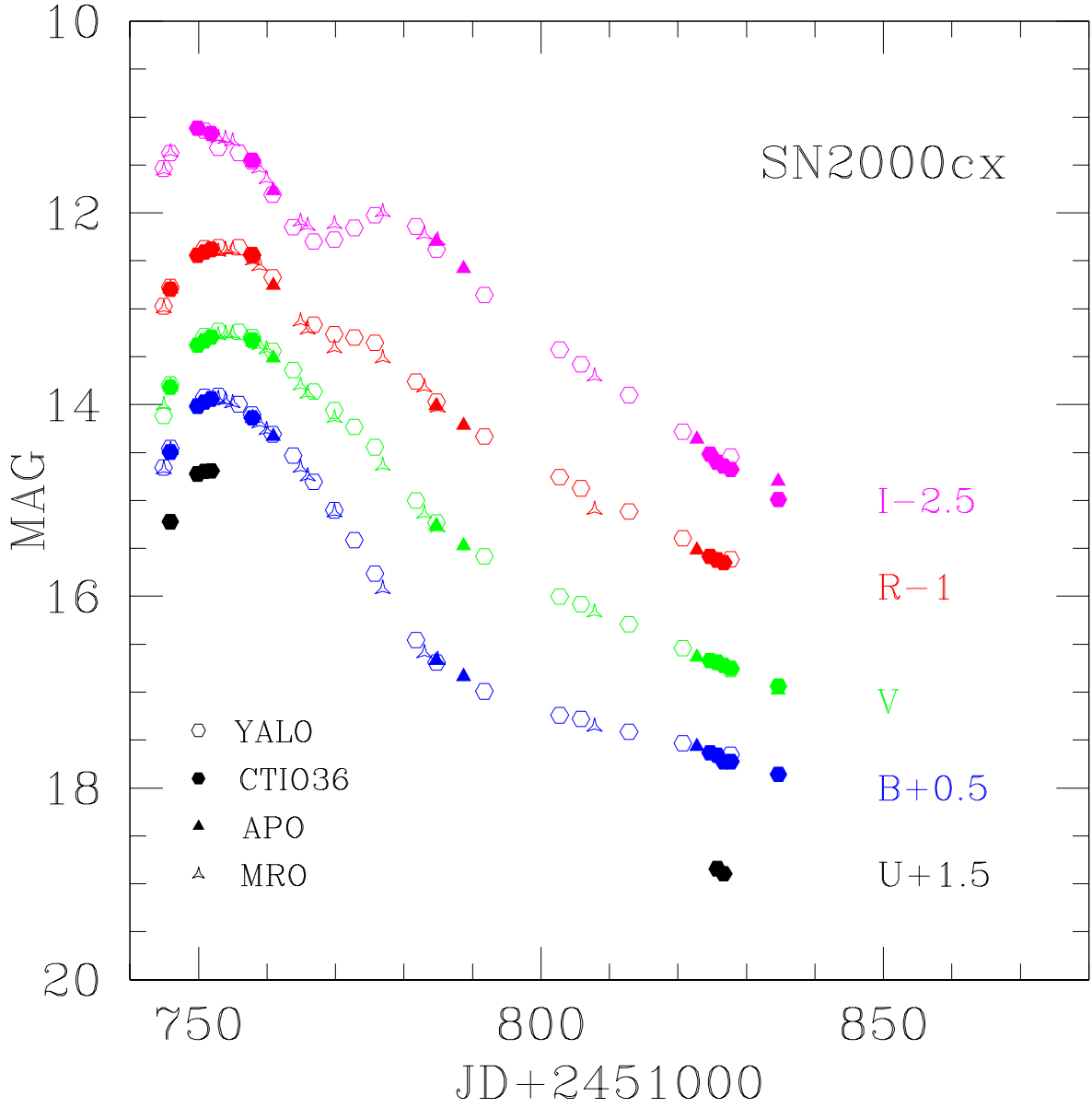}
\caption{$U$-, $B$-, $V$-, $R$-, and $I$-band light curves of SN 2000cx,
showing the optical data presented in this paper.
The data are coded by telescope.}
\end{figure}

\begin{figure}
\figurenum{3}
\epsscale{1}
\plotone{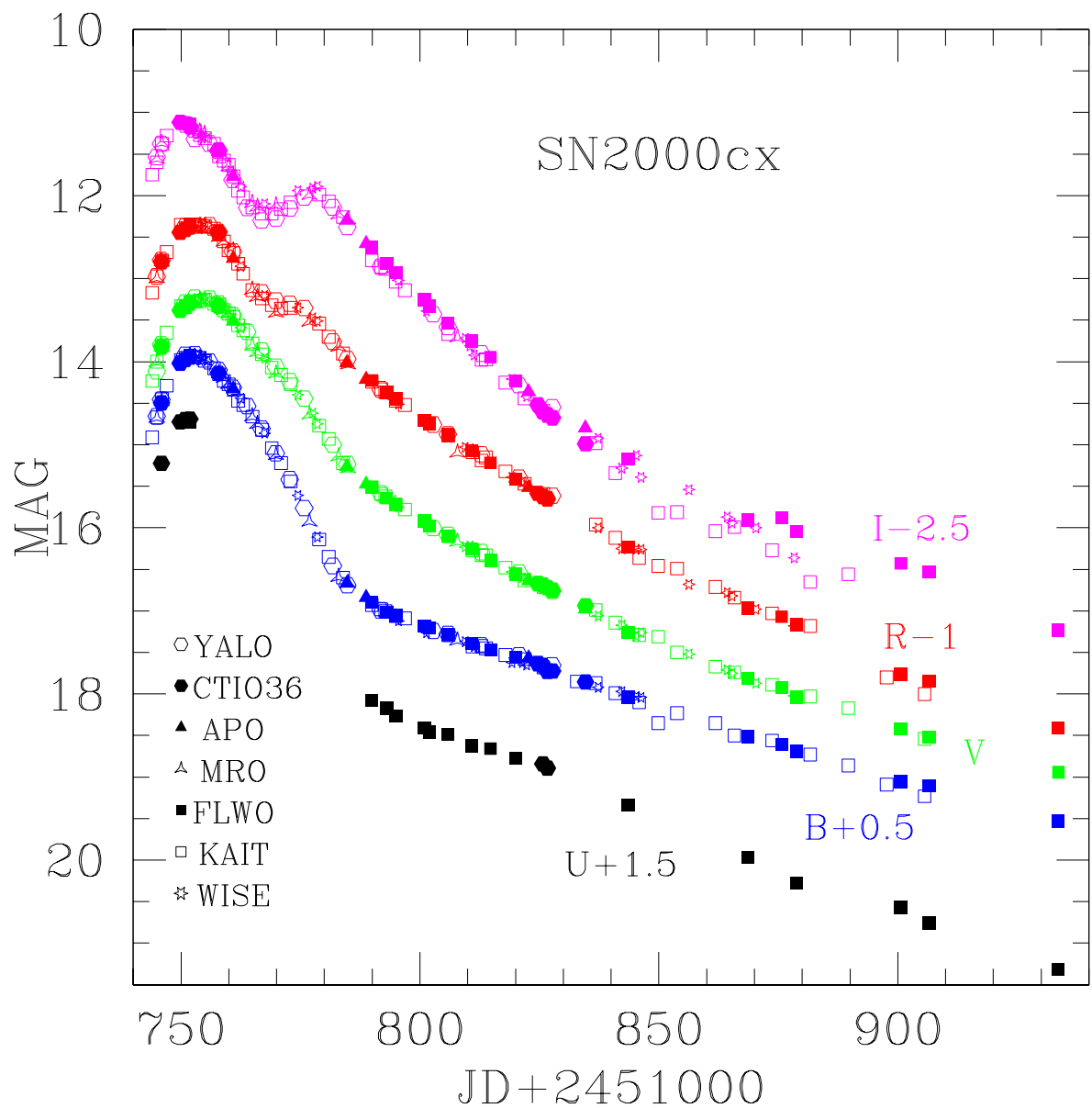}
\caption{Same as Fig. 2, but with the addition of KAIT and Wise Observatory
data given by Li et al. (2001), plus the Jha (2000) data obtained
at the Fred L. Whipple Observatory.}
\end{figure}

\begin{figure}
\figurenum{4}
\epsscale{1}
\plotone{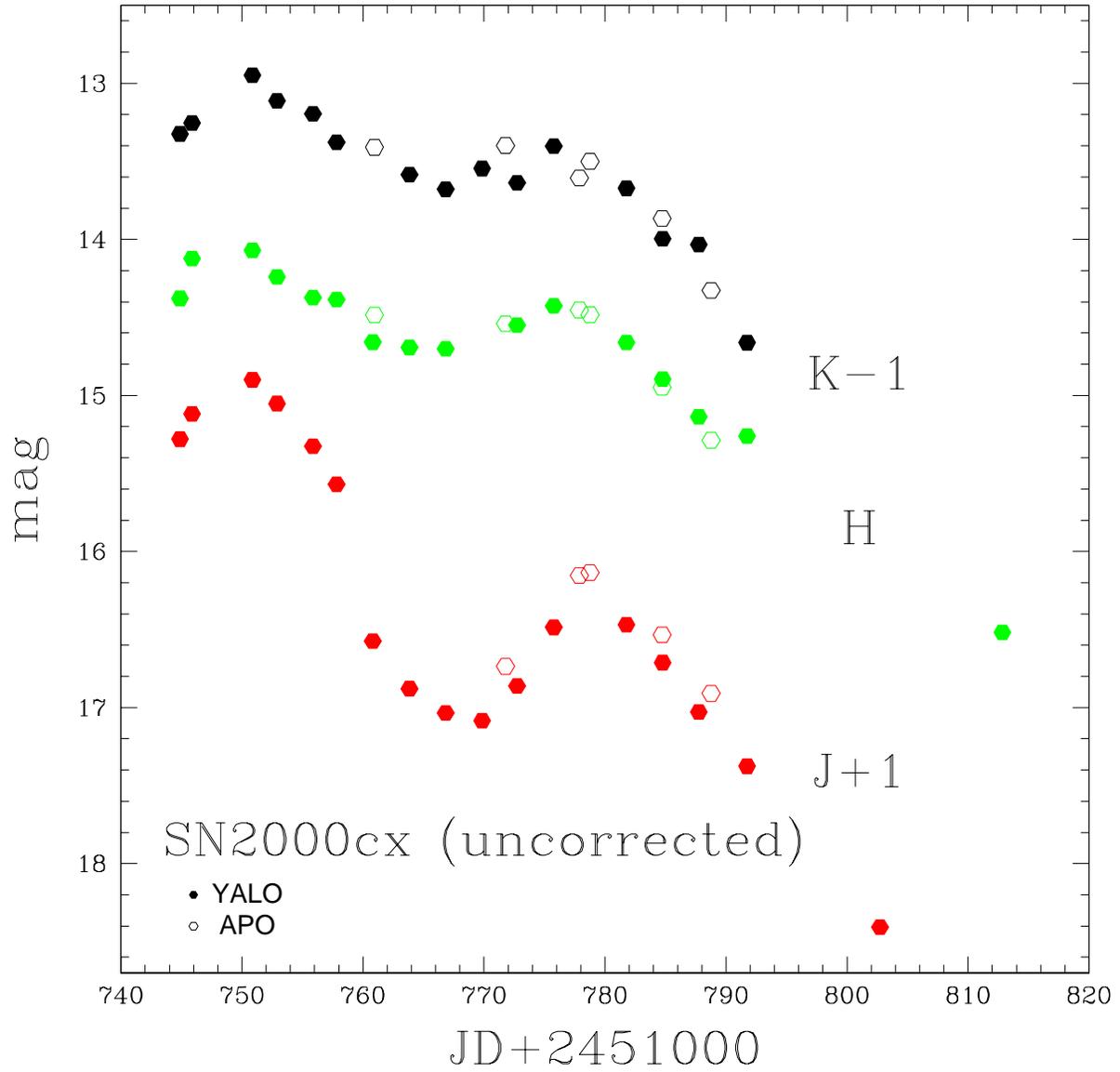}
\caption{Infrared light curves for YALO and APO. The $K$- and $J$-band
data have been offset by $-1$ and +1 magnitudes, respectively.
APO data for Julian Date 2,451,834 are off the right hand side of the plot.}
\end{figure}

\begin{figure}
\figurenum{5}
\epsscale{1}
\plotone{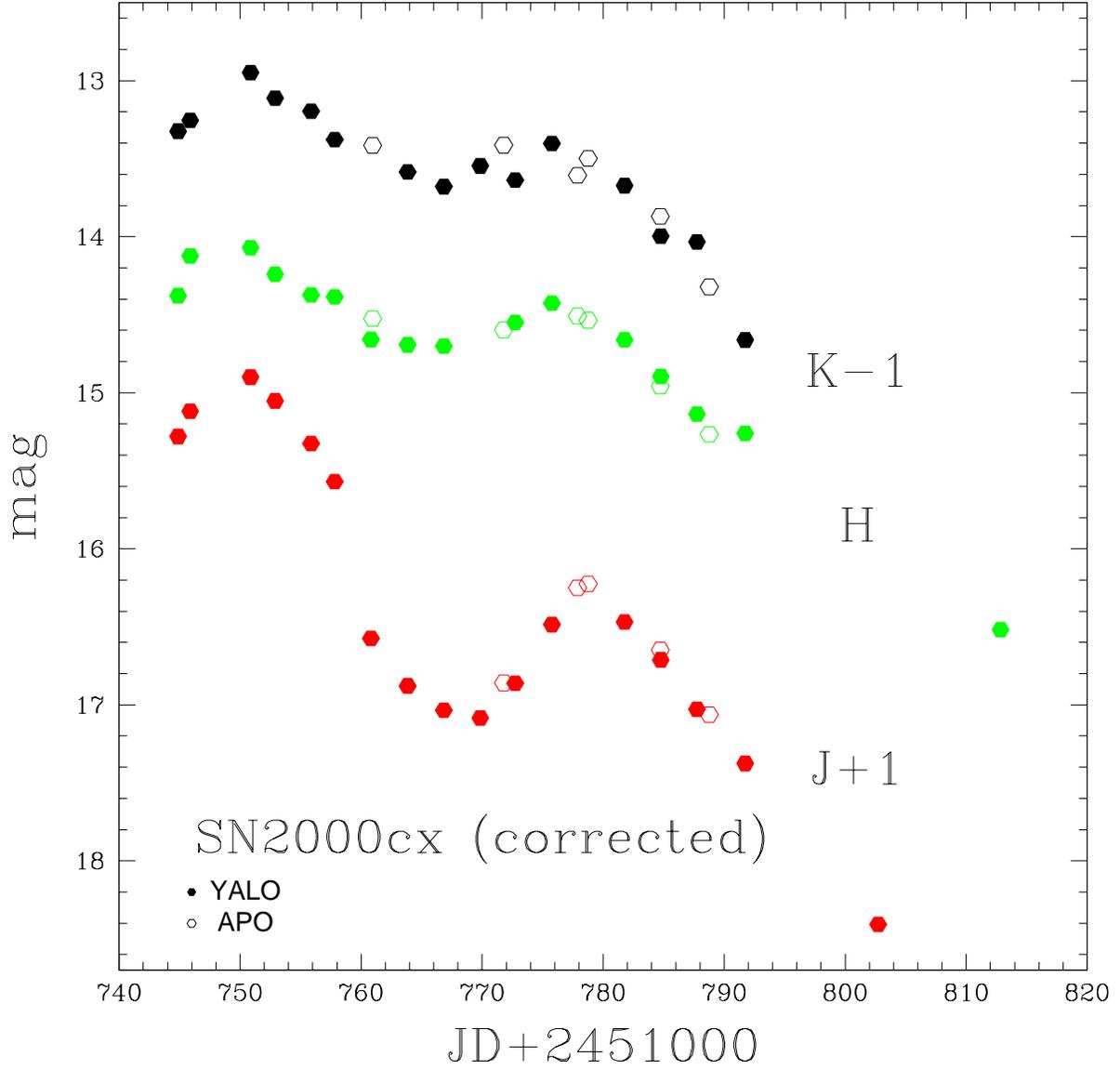}
\caption{Same as Fig. 4, but the APO data have been corrected by the 
values given in Table 9. }
\end{figure}

\begin{figure}
\figurenum{6}
\epsscale{0.8}
\plotone{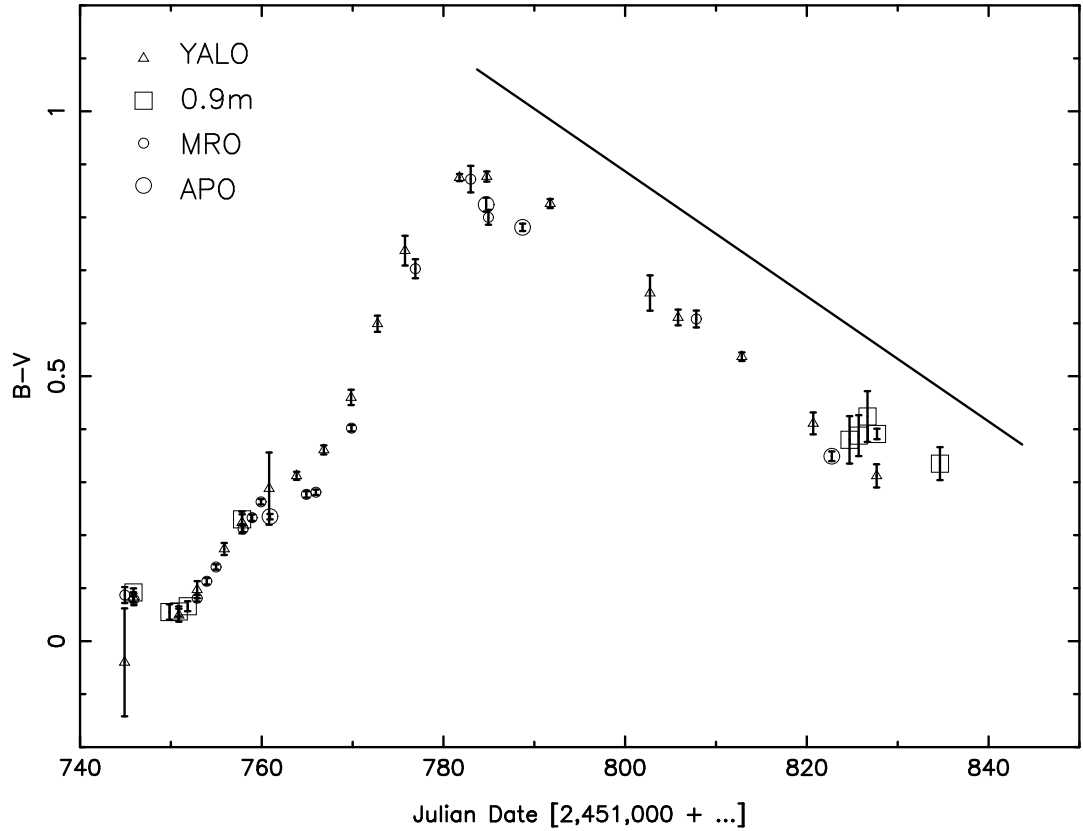}
\vspace {5 mm}
\caption{$B-V$ color curve of SN 2000cx, showing the optical data
presented in this paper.  E($B-V$) = 0.082 has been subtracted from
the data to eliminate the effect of dust in our Galaxy (Schlegel et al.
1998). The ``zero reddening line'' of Lira (1995) is shown, adjusted
to the time of $V$-band maximum given in Table 10. If SN 2000cx has
any host reddening, then the data points corresponding to unreddened
photometry would be even further below the Lira line.}
\end{figure}

\begin{figure}
\figurenum{7}
\epsscale{0.8}
\plotone{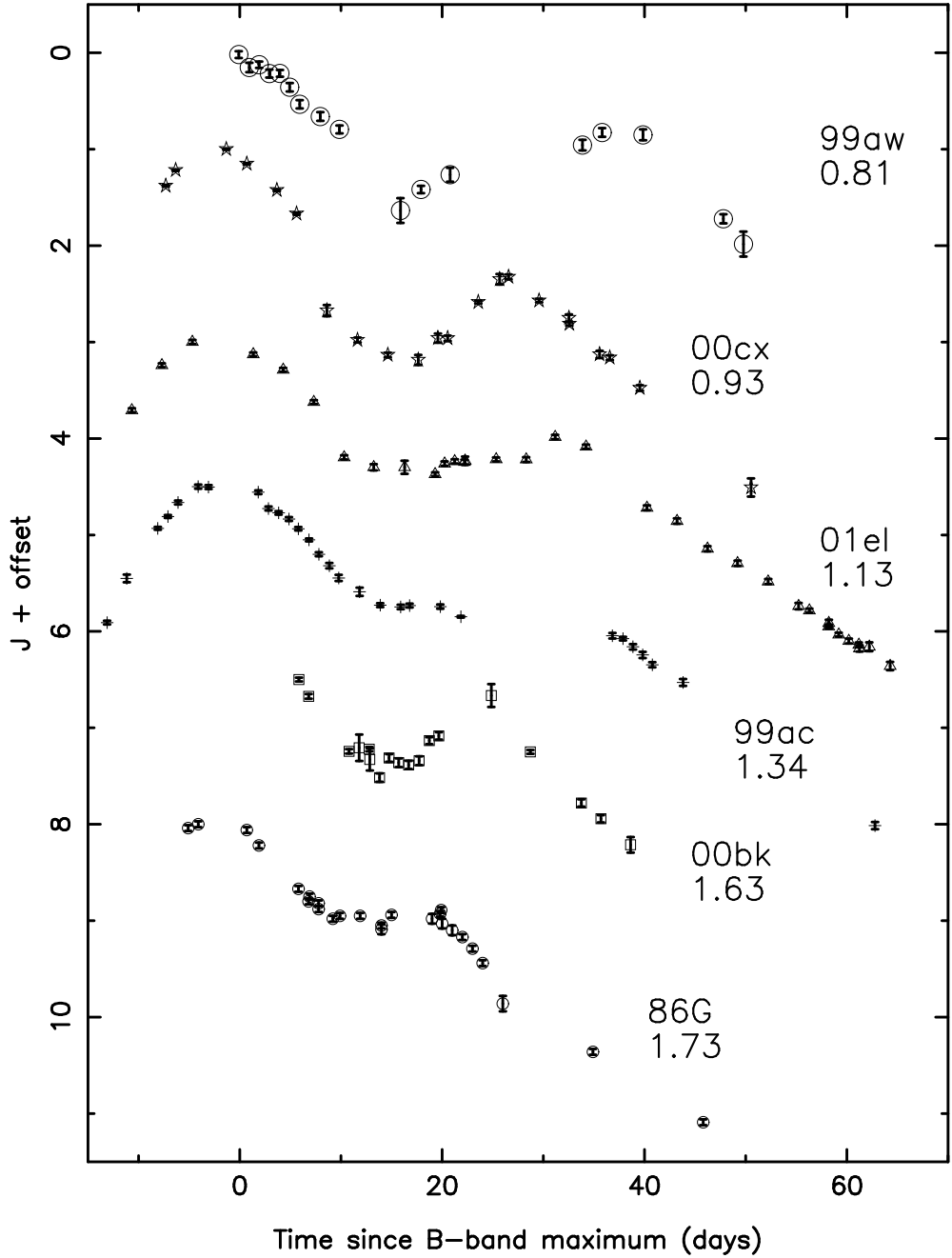}
\caption{$J$-band light curve of SN 2000cx along with data of SNe
1999aw (Strolger et al. 2002), 2001el (Krisciunas et al. 2003), 
1999ac (Phillips et al. 2002a, 2002b), 2000bk (Krisciunas et al. 2001),
and 1986G (Frogel et al. 1987). The light curves are ordered from top
to bottom by the decline rate parameter $\Delta$m$_{15}$($B$).}
\end{figure}

\begin{figure}
\figurenum{8}
\epsscale{0.8}
\plotone{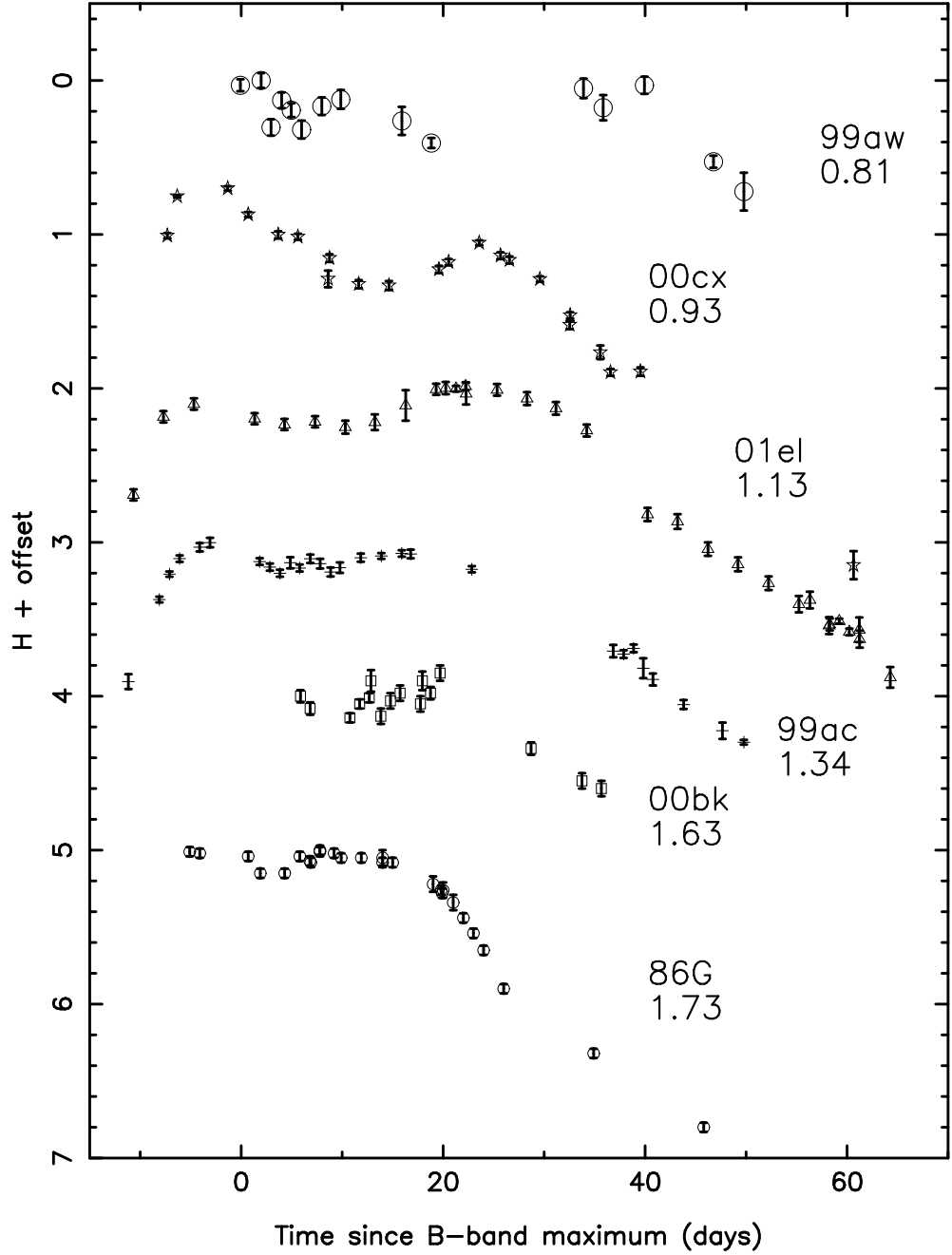}
\caption{Same as for Fig. 7, but for the $H$-band.}
\end{figure}

\begin{figure}
\figurenum{9}
\epsscale{0.8}
\plotone{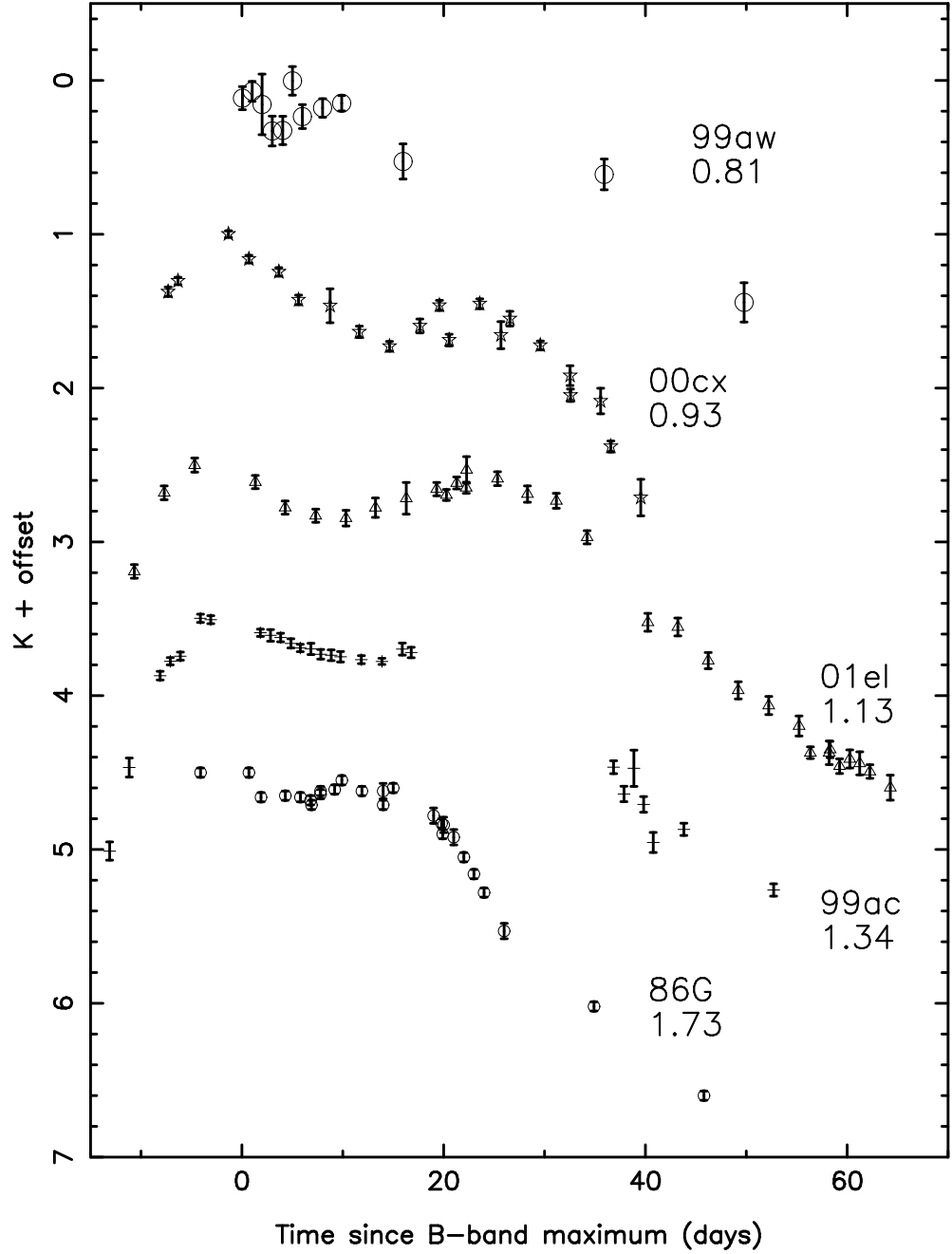}
\caption{Same as for Fig. 7, but for the $K$-band.  No $K$-band data
were taken of SN 2000bk.}
\end{figure}

\begin{figure}
\figurenum{10}
\epsscale{0.8}
\plotone{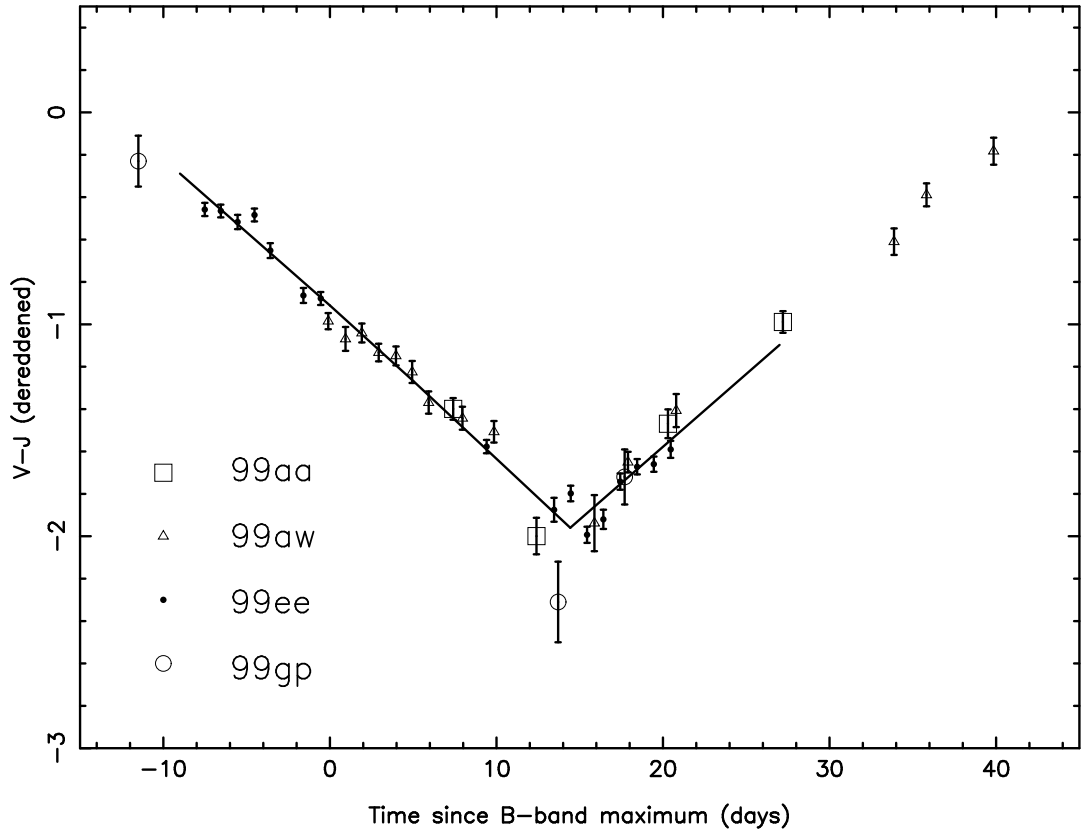}
\vspace {5 mm}
\caption{Dereddened $V-J$ colors of slowly declining Type Ia SNe.  SNe 1999aa,
1999aw, and 1999gp were assumed to be unreddened in their host galaxies.
For these three objects, only the reddening due to dust in our Galaxy has been
subtracted (Schlegel et al. 1998).  SN 1999ee has non-zero host reddening, which,
along with its Galactic reddening, has been subtracted for the purposes of this
graph.}
\end{figure}

\begin{figure}
\figurenum{11}
\epsscale{0.8}
\plotone{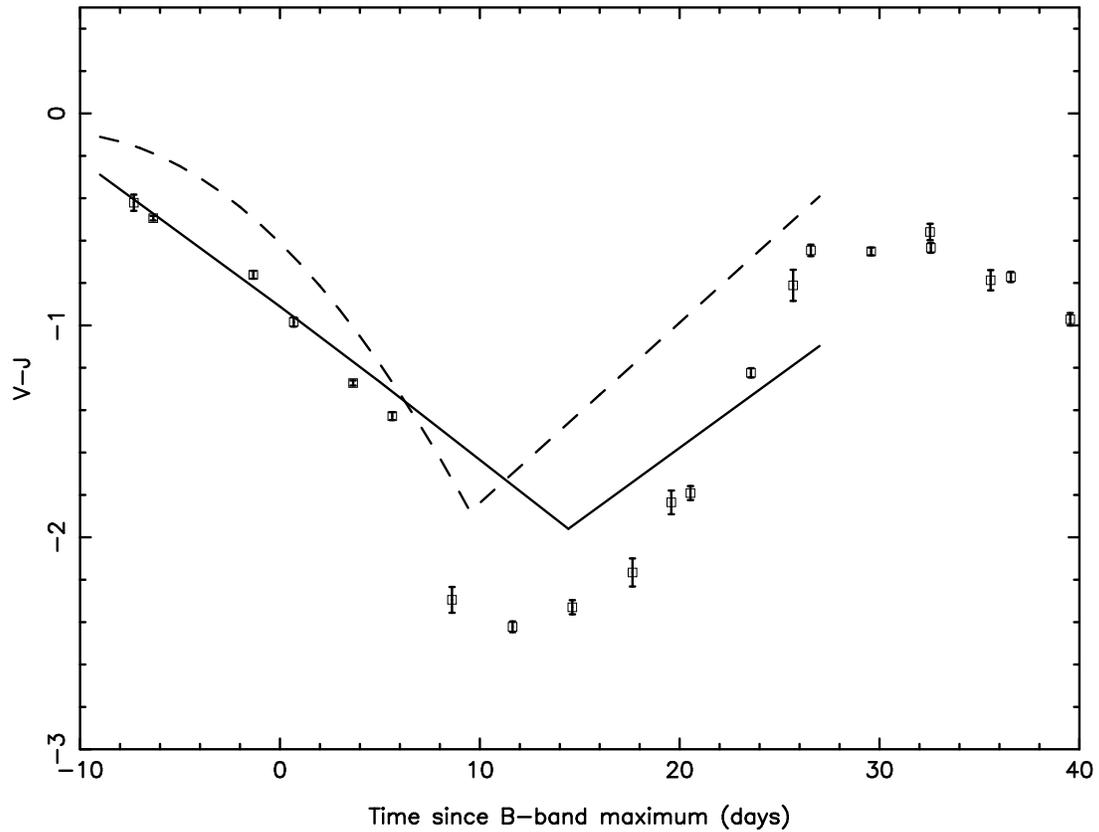}
\vspace {5 mm}
\caption{$V-J$ colors of SN 2000cx, with E($V-J$) = 0.182 subtracted
to account for the effect of dust in our Galaxy.  The dashed line is the
unreddened locus given by Krisciunas et al. (2000) for Type Ia SNe with
mid-range $B$-band decline rates.  The solid line is the unreddened locus
derived from photometry of four slow decliners (SNe 1999aa, 1999aw, 1999gp,
and 1999ee) shown in Fig. 10.}
\end{figure}

\begin{figure}
\figurenum{12}
\epsscale{0.8}
\plotone{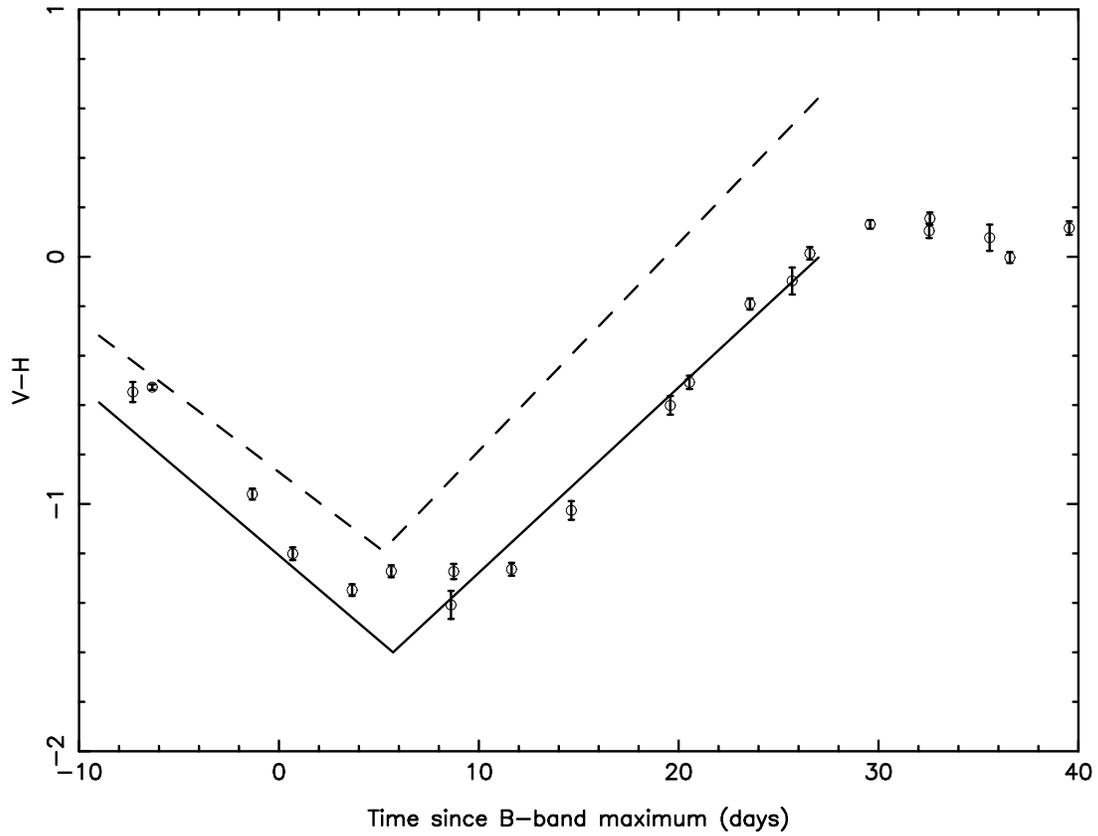}
\vspace {5 mm}
\caption{$V-H$ colors of SN 2000cx. The solid line is based on
data of the slowly declining SNe 1999aw, 1999gp, and 1999ee. The dashed line is based on
Type Ia SNe with mid-range $B$-band decline rates (Krisciunas et al. 2000).
To eliminate the effect of dust in our Galaxy on the colors, E($V-H$) = 
0.210 mag has been subtracted from the SN 2000cx data.}
\end{figure}

\begin{figure}
\figurenum{13}
\epsscale{0.8}
\plotone{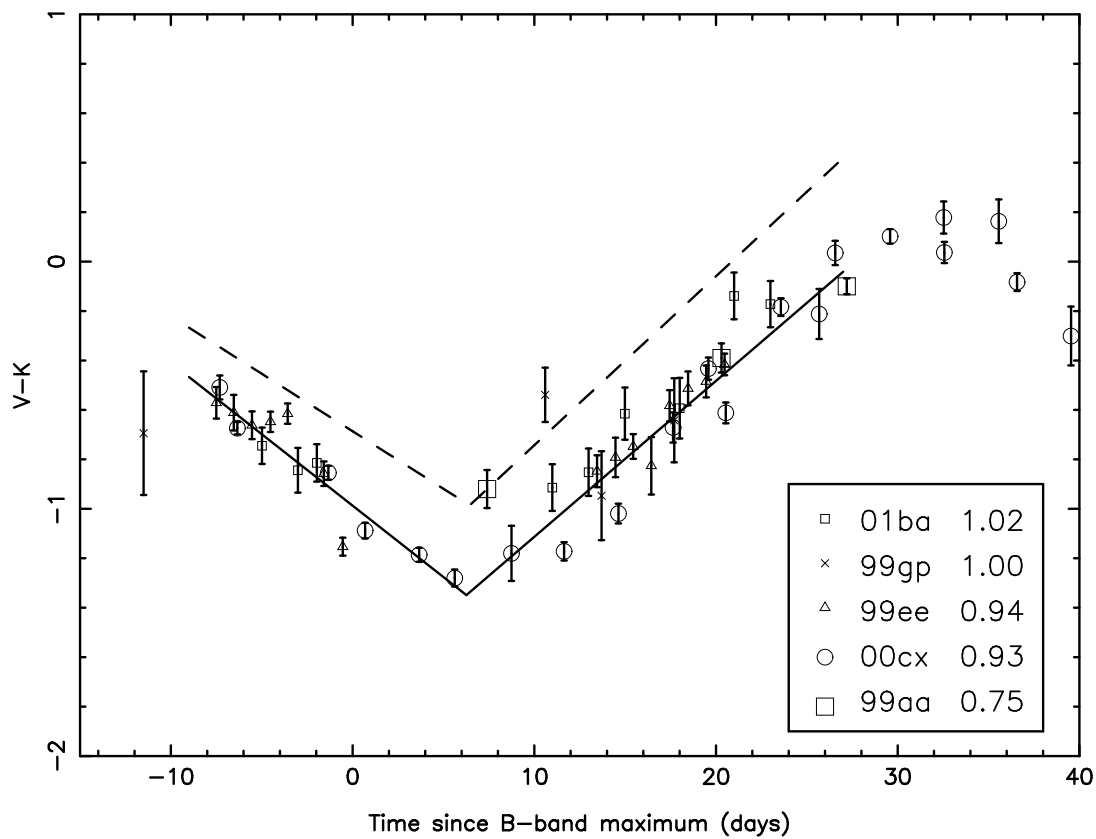}
\vspace {5 mm}
\caption{$V-K$ colors of SN 2000cx and four other slowly declining
Type Ia SNe.  The values of $\Delta$m$_{15}$($B$) are given in the box.
The data of SNe 2000cx, 1999gp, and 1999aa have only been
corrected for Galactic reddening (Schlegel et al. 1998).  Data of SNe 1999ee
and 2001ba have been corrected for host reddening and Galactic reddening.
The dashed line is based on dereddened Type Ia SNe 
with mid-range $B$-band decline 
rates (Krisciunas et al. 2000).  Clearly, the dereddened $V-K$ colors of the
objects shown here are bluer than the locus based on mid-range decliners.
From $10 < t < 21 $ d the SN~2000cx data are the bluest.}
\end{figure}

\begin{figure}
\figurenum{14}
\epsscale{1}
\plotone{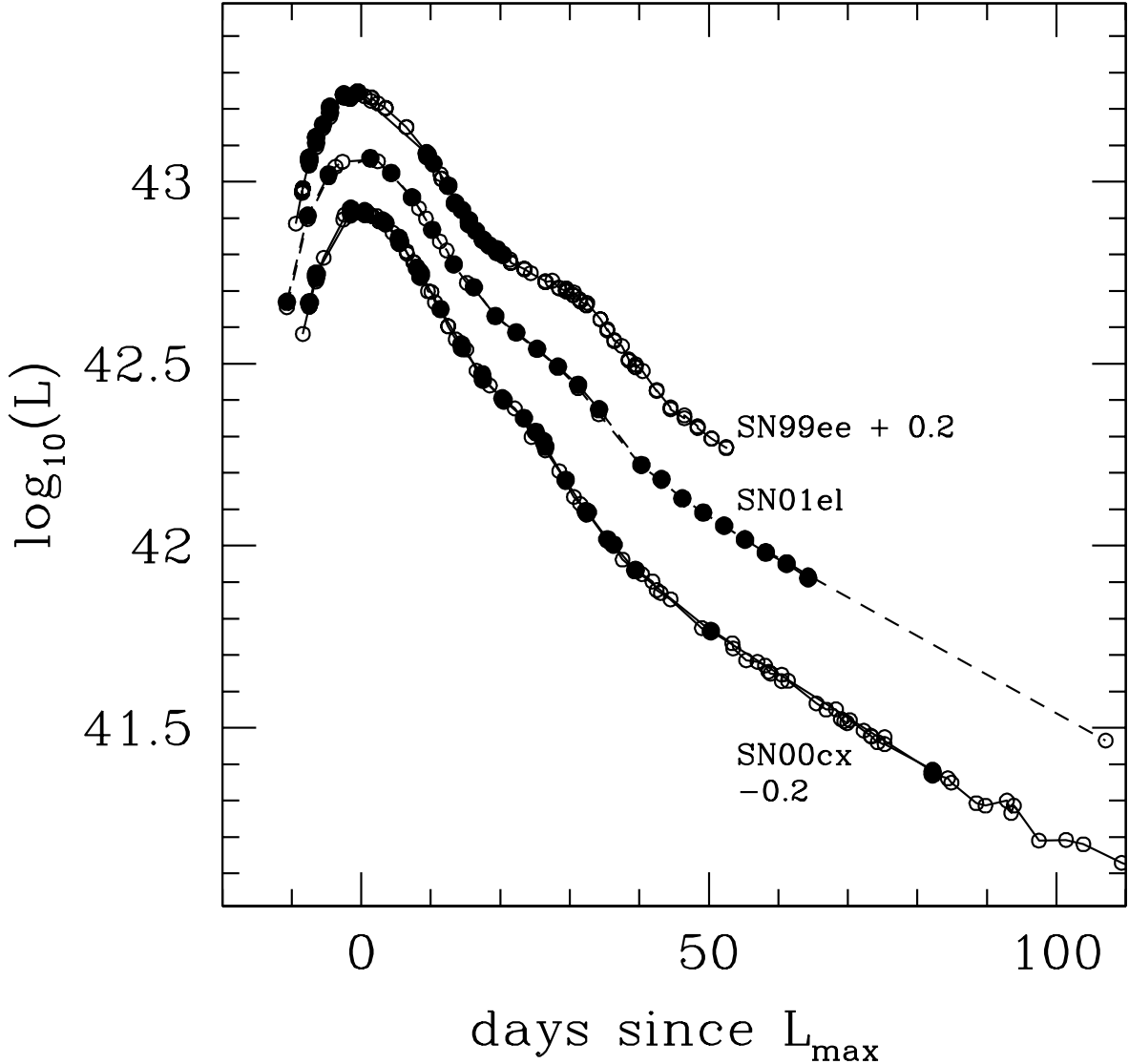}
\caption{''Uvoir'' bolometric light curves for SNe
1999ee, 2001el, and 2000cx plotted against the time from maximum
bolometric luminosity. The light curves for SN~1999ee and SN~2000cx have
been shifted by +0.2dex (99ee) and --0.2dex (00cx) for clarity. We have
assumed a distance modulus of 32.47 for SN~2000cx. The closed circles
are the $UBVRIJHK$ integrations and the open circles are the $UBVRI$
integration. Both integrations include extrapolations to account for
the missing flux outside of the integration limits. SN~1999ee and
SN~2001el are normal Type Ia SNe with $\Delta$m$_{15}$($B$) values of 0.94
and 1.13. Note the greatly reduced luminosity in the inflection point
around day 30.}
\end{figure}

\begin{figure}
\figurenum{15}
\epsscale{1}  
\plotone{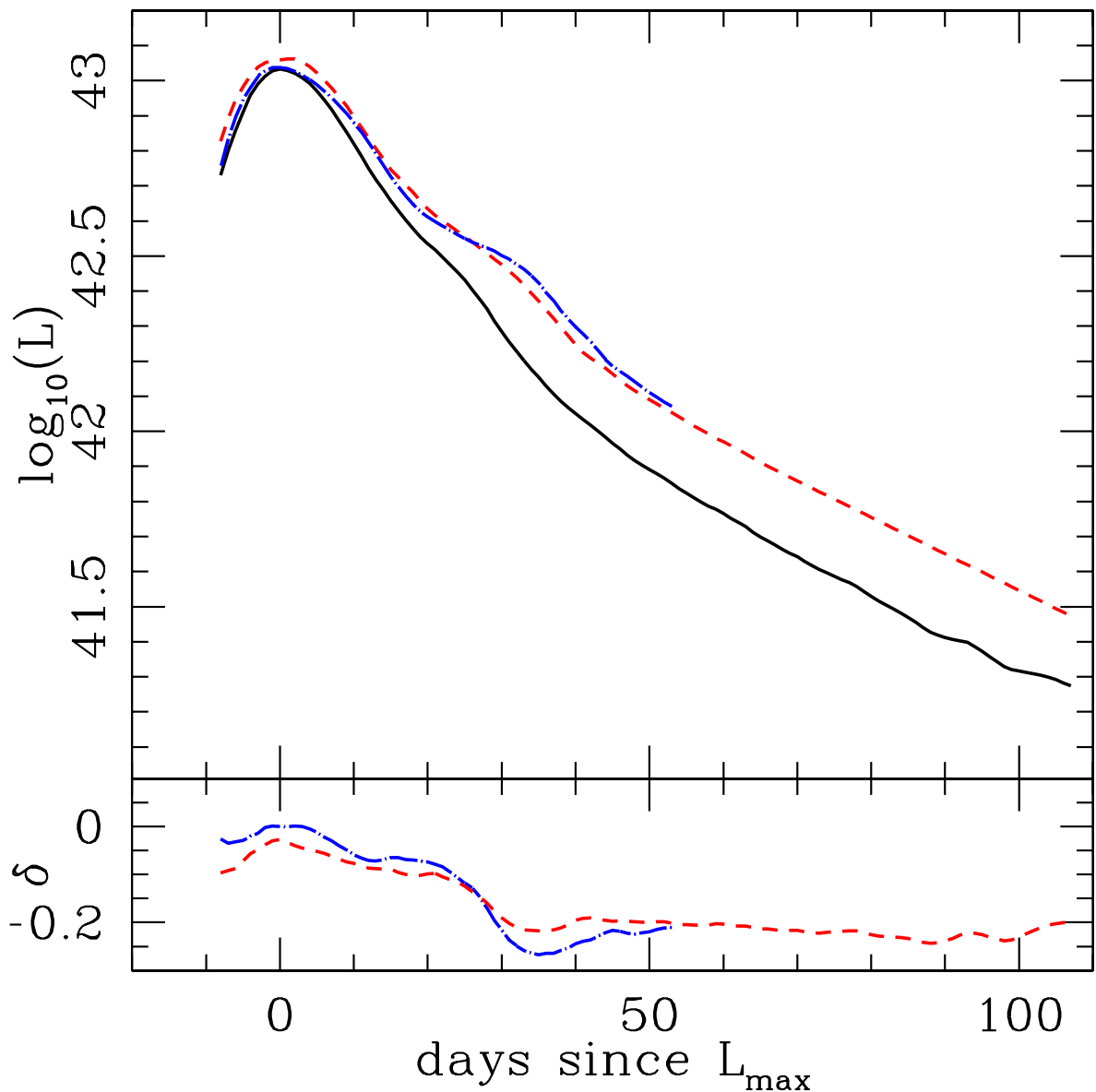}
\caption {Smoothed curve representations of the ``uvoir''
bolometric luminosity for SN~2000cx (solid curve), SN~1999ee (dot-dashed
curve), and SN~2001el (dashed curved). Note that the curves have not been
shifted as they were in Fig. 14. The bottom panel shows the
difference between the SN~2000cx bolometric light curve and the light
curves for SN~1999ee (dot-dashed curve) and SN~2001el (dashed curve). 
This plot shows there are two major differences between SN~2000cx and
these other ``normal'' Type Ia's: the lack of bolometric flux around
the time of the secondary $I$ maximum at 30 days, and the low
bolometric flux on the exponential tail after day 50 compared to the
peak luminosity.}
\end{figure}

\begin{figure}
\figurenum{16}
\epsscale{1}  
\plotone{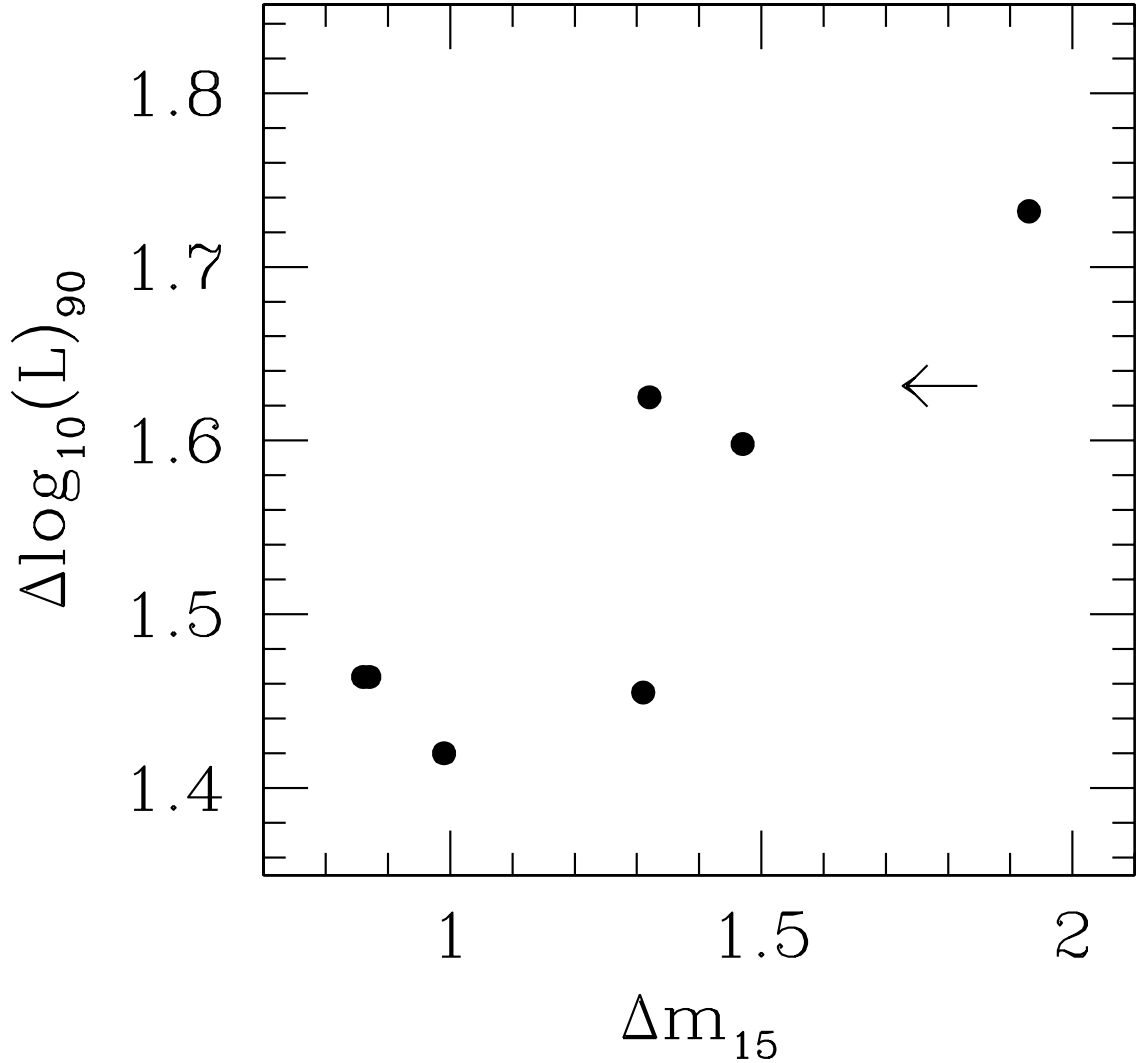}
\caption{The difference in the peak bolometric luminosity
compared to the luminosity 90 days after maximum light (in units of
dex) for the sample of supernovae in Figure 5 of Contardo et al. (2000) as
a function of $\Delta$m$_{15}$($B$). We mark the luminosity difference
observed for SN~2000cx with an arrow. Evidently intrinsically fainter
supernovae (those with larger $\Delta$m$_{15}$) have larger peak-to-
tail luminosities. The position of SN~2000cx would associate it with
the sub-luminous class of SNe.}
\end{figure}

\begin{figure}
\figurenum{17}
\epsscale{1}
\plotone{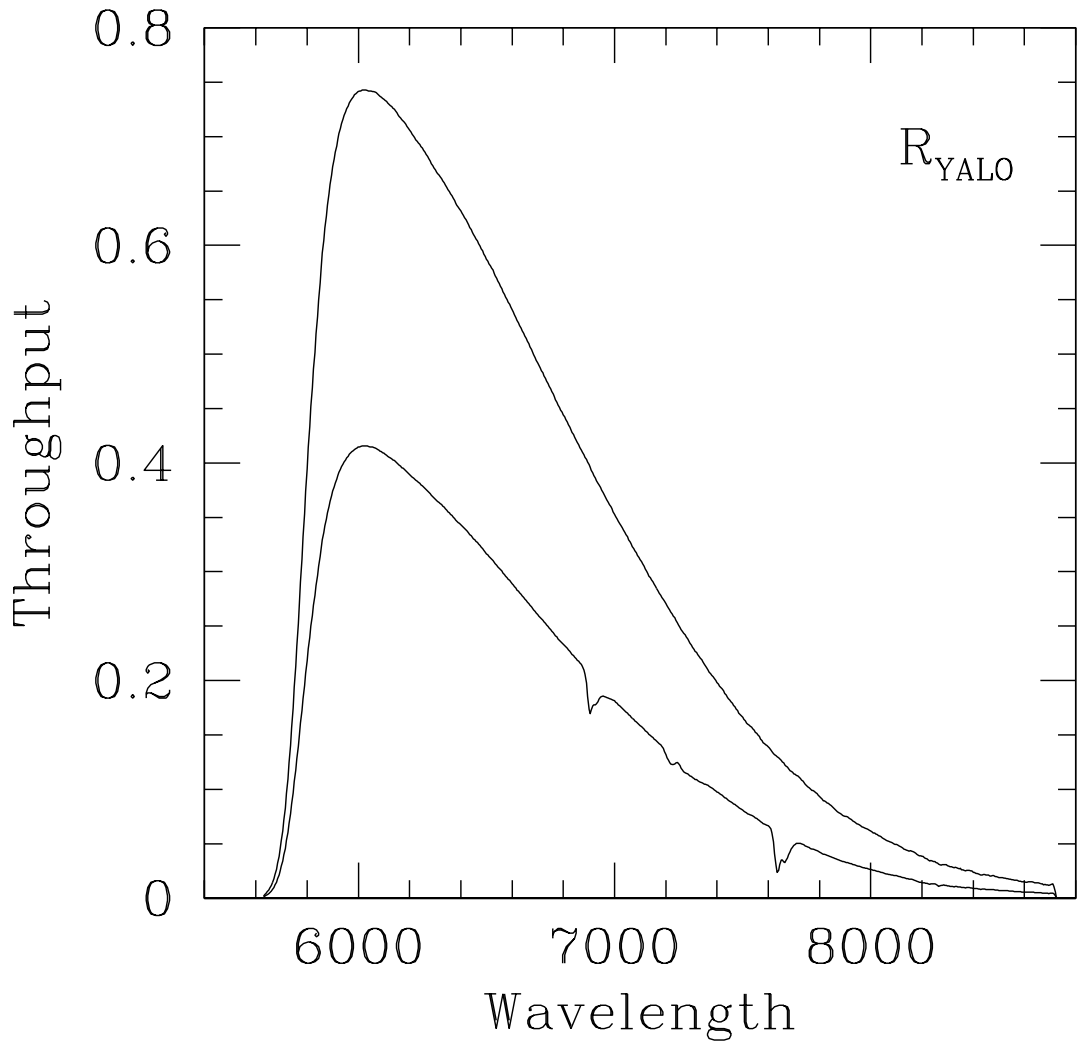}
\caption{Upper curve: the transmission function of the $R$ filter
used for the YALO observations reported here.  Lower curve:
the filter transmission function multiplied by an atmospheric
transmission function, quantum efficiency vs. wavelength, and
two aluminum reflections, giving the effective transmission with
the system for $R$.  Both curves have been shifted 280 \AA\ to
shorter wavelengths to account for the cooling of the filter in
the dewar and so that the color term obtained from synthetic
photometry matches that determined from observations of Landolt
(1992) standards.}
\end{figure}

\end{document}